\documentclass{elsart}
\pdfoutput=1
\usepackage{graphicx,url}
\usepackage{subfigure}
\usepackage{amsmath}
\usepackage{amsfonts}
\usepackage{amssymb}
\usepackage{subfigure}
\newcounter{bla}
\newenvironment{refnummer}{%
\list{[\arabic{bla}]}%
{\usecounter{bla}%
 \setlength{\itemindent}{0pt}%
 \setlength{\topsep}{0pt}%
 \setlength{\itemsep}{0pt}%
 \setlength{\labelsep}{2pt}%
 \setlength{\listparindent}{0pt}%
 \settowidth{\labelwidth}{[9]}%
 \setlength{\leftmargin}{\labelwidth}%
 \addtolength{\leftmargin}{\labelsep}%
 \setlength{\rightmargin}{0pt}}}
 {\endlist}

\newtheorem{defi}{Definition}
  
\newcommand{\bra}[1]{\left\langle{#1}\right\vert}
\newcommand{\ket}[1]{\left\vert{#1}\right\rangle}
\def\Hil{\mathcal{H}}
\def\L{\mathcal{L}}
\def\Z{\mathbb{Z}}
\def\C{\mathbb{C}}

\begin{document}
\begin{frontmatter}

\title{The QWalk Simulator of Quantum Walks}

\author[a]{F.L.~Marquezino\thanksref{author}},
\author[a]{R.~Portugal},

\thanks[author]{Corresponding author: franklin@lncc.br}

\address[a]{LNCC, Laborat\'orio Nacional de Computa\c c\~ao Cient\'{\i}fica\\Av. Get\'ulio Vargas~333, CEP~25651-075, Petr\'opolis, RJ, Brazil}

\begin{abstract}
  %Type your abstract here.
Several research groups are giving special attention to quantum walks recently, because this research area have been used with success in the development of new efficient quantum algorithms. A general simulator of quantum walks is very important for the development of this area, since it allows the researchers to focus on the mathematical and physical aspects of the research instead of deviating the efforts to the implementation of specific numerical simulations. In this paper we present QWalk, a quantum walk simulator for one- and two-dimensional lattices. Finite two-dimensional lattices with generic topologies can be used. Decoherence can be simulated by performing measurements or by breaking links of the lattice. We use examples to explain the usage of the software and to show some recent results of the literature that are easily reproduced by the simulator.

\begin{flushleft}
  %Insert your suggested PACS number here
PACS: 03.67.Lx, 05.40.Fb, 03.65.Yz

\end{flushleft}

\begin{keyword}
Quantum walk; quantum computing; Quantum Mechanics; double-slit; broken links
  % Please give some freely chosen keywords that we can use in a
  % cumulative keyword index.
\end{keyword}

\end{abstract}

\end{frontmatter}

% Computer program descriptions should contain the following
% PROGRAM SUMMARY.

{\bf PROGRAM SUMMARY}
  %Delete as appropriate.

\begin{small}
\noindent
{\em Manuscript Title:} The QWalk Simulator of Quantum Walks\\
{\em Authors:} F.L. Marquezino and R. Portugal                \\
{\em Program Title:} QWalk                                    \\
{\em Journal Reference:}                                      \\
  %Leave blank, supplied by Elsevier.
{\em Catalogue identifier:}                                   \\
  %Leave blank, supplied by Elsevier.
{\em Licensing provisions:} GNU General Public Licence.        \\
  %enter "none" if CPC non-profit use license is sufficient.
{\em Programming language:} C                                 \\
{\em Computer:} Any computer with a C compiler that accepts ISO C99 complex arithmetic (recent versions of GCC, for instance). Pre-compiled versions are also provided.\\
  %Computer(s) for which program has been designed.
{\em Operating system:} The software should run in any operating system with a recent C compiler. Successful tests were performed in Linux and Windows. \\
  %Operating system(s) for which program has been designed.
{\em RAM:} Less than 10~MB were required for a two-dimensional lattice of size $201\times 201$. About 400~MB, for a two-dimensional lattice of size $1601\times 1601$.\\
  %RAM in bytes required to execute program with typical data.
%{\em Number of processors used:} 
  %If more than one processor.
{\em Supplementary material:} Several examples of input files are provided. \\
  % Fill in if necessary, otherwise leave out.
{\em Keywords:} Quantum walk; quantum computing; Quantum Mechanics; simulation; double-slit; broken links; C.\\
  % Please give some freely chosen keywords that we can use in a
  % cumulative keyword index.
{\em PACS:} 03.67.Lx, 05.40.Fb, 03.65.Yz                    \\
  % see http://www.aip.org/pacs/pacs.html 
{\em Classification:}                                         \\
  %Classify using CPC Program Library Subject Index, see (
  % http://cpc.cs.qub.ac.uk/subjectIndex/SUBJECT_index.html)
  %e.g. 4.4 Feynman diagrams, 5 Computer Algebra.
%{\em External routines/libraries:}                                      \\
  % Fill in if necessary, otherwise leave out.
{\em Subprograms used:} The simulator generates gnuplot scripts that can be used to make graphics of the output data. \\
  %Fill in if necessary, otherwise leave out.

{\em Nature of problem:} Classical simulation of discrete quantum walks in one- and two-dimensional lattices.\\
  %Describe the nature of the problem here.
   \\
{\em Solution method:} Iterative approach without explicit representation of evolution operator.\\
  %Describe the method solution here.
   \\
{\em Restrictions:} The available amount of RAM memory imposes a limit on the size of the simulations.\\
  %Describe any restrictions on the complexity of the problem here.
   \\
{\em Unusual features:} The software provides an easy way of simulating decoherence through detectors or random broken links. In the two-dimensional simulations it also allows the definition of permanent broken links, besides calculation of total variation distance (from the uniform and from an approximate stationary distribution) and the choice between two different physical lattices. It also provides an easy way of performing measurements on specific sites of the 2D lattice and the analysis of observation screens. In one-dimensional simulations it allows the choice between three different lattices. Both one- and two-dimensional simulations facilitates the generation of graphics by automatically generating gnuplot scrips.\\
  %Describe any unusual features of the program/problem here.
   \\
{\em Additional comments:} An earlier version of QWalk was first presented in~[1].\\
  %Provide any additional comments here.
   \\
{\em Running time:} The simulation of $100$ steps for a two-dimensional lattice of size $201\times 201$ took less than 2 seconds on a Pentium IV 2.6GHz with 512MB of RAM memory, 512KB of cache memory and under Linux. It also took about 15 minutes for a lattice of size $1601\times 1601$ on the same computer. Optimization option \verb|-O2| was used during compilation for these tests.\\
  %Give an indication of the typical running time here.
   \\
{\em References:}
\begin{refnummer}
\item Marquezino, F.L. and Portugal, R., QWalk: Simulador de Caminhadas Qu\^anticas, in \emph{Proceedings of 2nd WECIQ}, pages 123-132, Campina Grande, Brazil, 2007, IQuanta.
       % This is the reference list of the Program Summary
       % Type references in text as [1], [2], etc.
       % This list is different from the bibliography, which
       % you can use in the Long Write-Up.
\end{refnummer}

\end{small}

\newpage

% In program descriptions the main text of the paper is listed under
% the heading LONG WRITE-UP.

\hspace{1pc}
{\bf LONG WRITE-UP}

\section{Introduction}
 
A very successful approach used to solve some problems in classical computing consists in using algorithms based on random walks. In fact, many computational problems are solved more efficiently by this technique~\cite{MotwaniR95}. In the 1990s the model of discrete quantum walks has been developed by Aharonov \emph{et al}~\cite{DavidovichAZ93} and the continuous model has been developed by Farhi and Gutmann~\cite{FarhiGutmann98}. The good results obtained with random walks in classical computing motivate us to investigate the applications of quantum walks to the development of new efficient quantum algorithms. 

Quantum algorithms based on quantum walks have already been developed with great success. Some examples are the search algorithm by Shenvi \emph{et al}~\cite{ShenviKW03} and the algorithm for element distinctness by Ambainis~\cite{Ambainis04}, both using the discrete-time model. Kempe~\cite{Kempe03} showed that a discrete-time quantum walker can cross a hypercube of dimension $d\geq 3$ exponentially faster than a classical random walk. Farhi \emph{et al}~\cite{FarhiNAND} have developed an efficient quantum algorithm for the Hamiltonian NAND tree based on the continuous-time quantum walk. 

Together with amplitude amplification and Fourier transform, quantum walks are one of the strategies for developing efficient quantum algorithms. Algorithms based on quantum walks are better than the Grover algorithm in some cases~\cite{Ambainis04}. Apart from the applications on the development of efficient quantum algorithms, which are important from a Computer Science perspective, the quantum walks have also properties that justify their investigation from the point of view of Physics. 
    
A general simulator of quantum walks is important for the development of this research area. Without such a simulator, the effort of the researchers are deviated to the implementation of specific numerical simulations while it should the focused in the physical and mathematical aspects of the research. The QWalk simulator allows the scientific community to perform important simulations on quantum walks with simple commands and even facilitates the generation of plots to visualize the results. In its present form, the QWalk simulator can reproduce most of the simulations present in research papers.

In Sec.~\ref{sec:walks} we briefly review the discrete quantum walks and some related definitions. In Sec.~\ref{sec:using} we present the QWalk simulator, using examples from recent results of literature in order to explain its usage. In Appendix~\ref{sec:further} we describe some options of the simulator that were not used in the examples.

\section{Quantum walks}
\label{sec:walks}

In the discrete one-dimensional quantum walk we consider a free particle---a walker---moving at each time step along a one-dimensional lattice. The direction of movement is given by an additional degree of freedom, the chirality of the particle, which can take two values, analogous to the result of the coin tossing in the classical random walk. In the two-dimensional quantum walk, similarly, the walker moves at each time step along a two-dimensional lattice. In this case two additional degrees of freedom are necessary, so that one may decide among four kinds of movements. In this Section we describe in details only the two-dimensional case. The one-dimensional walk, nevertheless, can be seen as a simple particularization of the case here exposed.

The Hilbert space considered in the two-dimensional walk is $\Hil_2 \otimes \Hil_2 \otimes \Hil_\infty$, where $\Hil_2\otimes \Hil_2$ is the Hilbert subspace associated with the chirality of the walker, i.e., the coin, and $\Hil_\infty$ is the Hilbert space associated with the position of the particle over the lattice. The basis for the coin subspace is $\mathcal{B}_C = \{\ket{j,k}: j,k\in \{0,1\} \}$ and the basis for the position subspace is $\mathcal{B}_S = \{\ket{m,n}: m,n\in \Z\}$.

The generic state of the quantum walker at time $t$ is
\begin{equation}
\ket{\Psi(t)} = \sum_{j,k=0}^{1}\sum_{m,n=-\infty}^{\infty}{\psi_{j,k;m,n}(t)\ket{j,k}\ket{m,n}},
\label{eq:state}
\end{equation}
with $\psi_{j,k;m,n}(t) \in \C$ and $\sum_{j,k}\sum_{m,n}{|\psi_{j,k;m,n}(t)|^2}=1$.
The evolution of the system over time is given by a unitary operator $U=S\circ (C\otimes I_P)$, where $S$ is the shift operator, $I_P$ is the identity operator which acts on the position subspace and $C$ is the coin operator, which acts on the $\Hil_2\otimes \Hil_2$ subspace. 

In this paper we address two different shift operators for the two-dimensional walk. The first of them has already been described in~\cite{AmandaPhysRev06} and is given by
\begin{equation}
S_a=\sum_{j,k=0}^{1}\sum_{m,n=-\infty}^{+\infty}{\ket{j,k}\bra{j,k}\otimes
\ket{m+(-1)^j,n+(-1)^k}\bra{m,n}}.
\label{eq:shiftA}
\end{equation}
One may notice that according to this shift operator the walker always moves along the diagonals of the mathematical lattice---it moves along the main diagonal if the value of the coin is either $\ket{00}$ or $\ket{11}$, and along the secondary diagonal if the value of the coin is either $\ket{01}$ or $\ket{10}$. In this case we say that the physical lattice is diagonal (in relation to the mathematical lattice).

We may also consider the possibility that, at time $t$, a link from site $(m,n)$ to one of its neighbours is broken. Then we need the functions
\begin{equation}
\L_1(j,k;m,n) = \begin{cases}
          (-1)^j&\mbox{if link to site } m+(-1)^j,n+(-1)^k \mbox{ is closed,}\\
	  0&\mbox{if link is open},
          \end{cases}
\end{equation}
and
\begin{equation}
\L_2(j,k;m,n) = \begin{cases}
          (-1)^k&\mbox{if link to site } m+(-1)^j,n+(-1)^k \mbox{ is closed,}\\
	  0&\mbox{if link is open},
          \end{cases}
\end{equation}
with $j,k\in \{0,1\}$. Whenever ${\L_1(1-j,1-k;m+(-1)^j,n+(-1)^k)=0}$ we must impose ${\L_1(j,k;m,n)=0}$, and similarly for $\L_2$. The technique of broken links for quantum walks was developed originally by Romanelli \emph{et al}~\cite{Abal+05} and later generalized for the two-dimensional case by Oliveira \emph{et al}~\cite{AmandaPhysRev06}.

\begin{figure}
\centering
\subfigure[Diagonal lattice]{\includegraphics[height=.35\textwidth, angle=270]{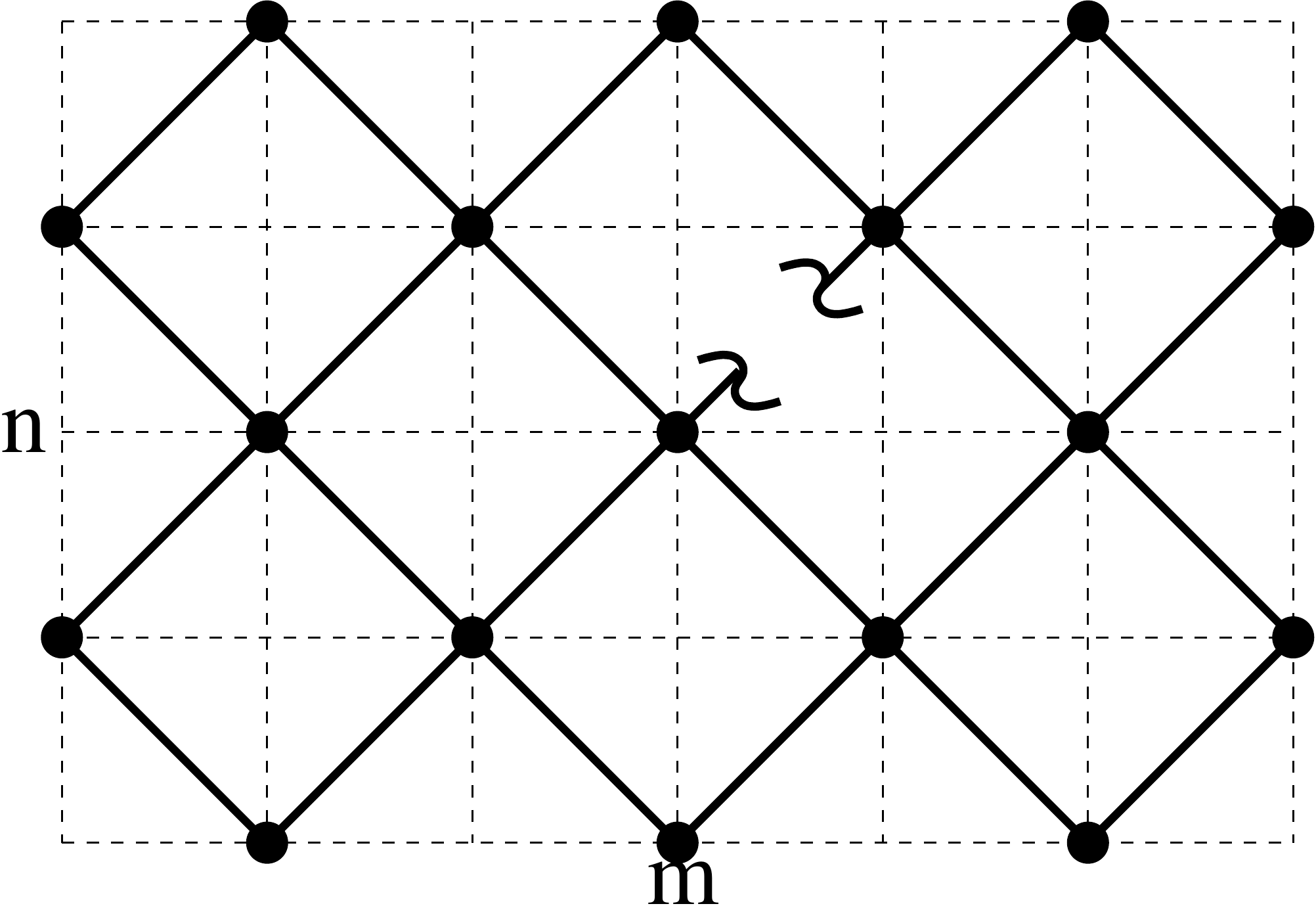}\label{fig:links}}
\subfigure[Natural lattice]{\includegraphics[height=.35\textwidth, angle=270]{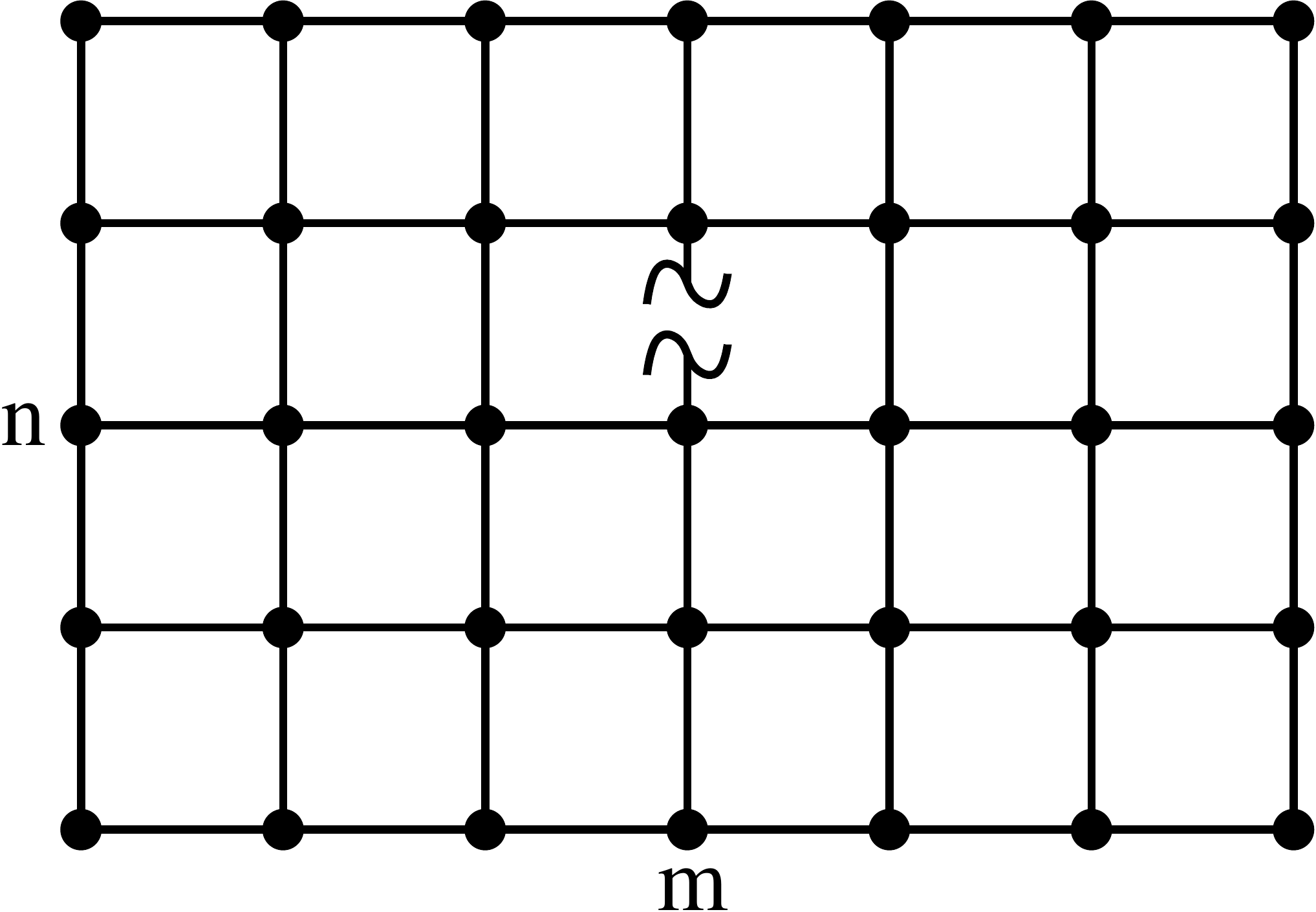}\label{fig:linksb}}
\caption{Part of the lattice for a two-dimensional quantum walk, showing a broken link.}
\end{figure}

In Fig.~\ref{fig:links} we have part of the lattice used in the two-dimensional quantum walk, with shift operator $S_a$. The  mathematical lattice is depicted by a dashed line and the physical lattice is depicted by a full line. In the example we have a broken link between sites $(m,n)$ and $(m+1,n+1)$.

If we apply the evolution operator $U_a$ to state~\eqref{eq:state} and include functions~$\L_1$ and~$\L_2$ as in~\cite{AmandaPhysRev06}, we obtain the evolution equation
\begin{multline}
\psi_{1-j,1-k;m,n}(t+1) =\\ \sum_{j^\prime, k^\prime=0}^{1}{C_{j+\L_1(j,k;m,n),k+\L_2(j,k;m,n);j^\prime,k^\prime}\psi_{j^\prime,k^\prime;m+\L_1(j,k;m,n),n+\L_2(j,k;m,n)}(t)}.
\end{multline}

We can also define a second shift operator in order to obtain a physical lattice that coincides with the mathematical lattice. This operator is given by
\begin{equation}
S_b=\sum_{j,d=0}^{1}\sum_{m,n=-\infty}^{+\infty}{\ket{j,d}\bra{j,d}\otimes
\ket{m+(-1)^j(1-\delta_{j,d}),n+(-1)^j\delta_{j,d}}\bra{m,n}}.
\label{eq:shiftB}
\end{equation}
We will observe through the examples of Section~\ref{sec:using} that the probability distributions obtained with operator $S_b$ differ from those obtained with operator $S_a$ only by a rotation of $\pi/4$. It will also be clear later on that, altough the movement described by $S_b$ may be more intuitive, operator $S_a$ has also some advantages.

If we intend to include the possibility of broken links in this second lattice, we need the function
\begin{equation}
\L(j,d;m,n) = \begin{cases}
          (-1)^j & \mbox{if link to } m+(-1)^j(1-\delta_{j,d}),n+(-1)^j\delta_{j,d} \mbox{ is closed,}\\
	  0&\mbox{if link is open},
          \end{cases}
\end{equation}
with $j,d\in \{0,1\}$. As in the previous case, we must impose ${\L(j,d;m,n)=0}$ whenever ${\L(1-j,1-d;m+(-1)^j(1-\delta_{j,d}),n+(-1)^j\delta_{j,d})=0}$. 

If we apply the evolution operator $U_b$ to state~\eqref{eq:state} and include function~$\L$, we obtain the evolution equation
\begin{multline}
\psi_{1-j,1-d;m,n}(t+1) =\\ \sum_{j^\prime, d^\prime=0}^{1}{C_{j+\L(j,d;m,n),d\oplus\L(j,d;m,n);j^\prime,d^\prime}\psi_{j^\prime,d^\prime;m+\L(j,d;m,n)(1-\delta_{j,d}),n+\L(j,d;m,n)\delta_{j,d}}(t)},
\end{multline}
where $\oplus$ is addition modulo $2$.

In Fig.~\ref{fig:linksb} we have part of the lattice used in the two-dimensional quantum walk with shift operator $S_b$. In this example we have a broken link between sites $(m,n)$ and $(m,n+1)$.

For the one-dimensional simulations, Qwalk provides three kinds of shift operators, which defines three different lattices: the infinite line, the finite one-dimensional lattice with reflecting boundaries and the cycle~\cite{Kendon06,KendonTregenna03}.

\subsection{Average distribution and mixing time}

By suitably defining permanent broken links we may study the quantum walk on a finite lattice with interesting topologies, such as a square box with reflecting boundaries~\cite{AmandaWeciq06}. Let 
\begin{equation}
P(m,n,T) = \sum_{j,k=0}^{1}{|\psi_{j,k;m,n}(T)|^2}
\end{equation}
be the probability of finding the walker on the site $(m,n)$ of the lattice at time $T$. In a classical random walk on a box this probability distribution converges to a stationary distribution. On the other hand, since the quantum walk must preserve unitarity, it does not present this convergence property. However, defining the average probability distribution as $\bar P(m,n,T) = \frac{1}{T}\sum_{t=0}^{T-1}{P(m,n,t)}$, Aharonov \emph{et al}~\cite{Aharonov+01} have proved that $\bar P(m,n,T)$ converges in the limit $T\rightarrow \infty$. Hence, define
\begin{equation}
\pi(m,n) = \lim_{T\rightarrow \infty}{\bar P(m,n,T)}.
\end{equation}
The rate at which the probability distribution approaches the limiting distribution $\pi(x)$ is captured by
\begin{defi}[Mixing time]
The mixing time $M_\epsilon$ of a quantum Markov chain is
$$M_\epsilon = \min\{ T | \forall t\geq T, \|\bar P_t - \pi \| \leq \epsilon \},$$
where $\|A-B\| = \sum_{m,n}|A(m,n)-B(m,n)|$ denotes the total variation distance.
\end{defi}
Usually $M_\epsilon$ depends on the initial conditions.

\section{Using the software}
\label{sec:using}

The QWalk simulator is quite easy to install on a Linux-like environment. One just needs to download the source code,\footnote{http://qubit.lncc.br/qwalk} uncompress it in any directory and finally use the command \verb|make|. The source code of the simulator was also successfully compiled in the Microsoft Windows operating system, using Dev-C++ 4.9.9.2 compiler, which can be downloaded for free in Internet\footnote{http://www.bloodshed.net.}. Detailed information on how to compile the simulator, as well as pre-compiled versions of it, are also provided in the web-site. Together with the downloaded files the user with knowledge in C programming also finds information on how to change the source code.

The simulator consists of three tools: \emph{qw1d} simulates quantum walks in one-dimensional lattices; \emph{qw2d}, in two-dimensional lattices; and \emph{qwamplify} improves the visualization of the plots generated by \emph{qw2d} by amplifying some regions. In this Section we give some examples from recent results of literature in order to show how to use the simulator. Other examples are also provided with the downloaded files.

\subsection{Simulations in two-dimensional lattices}

In order to use \emph{qw2d} it is necessary to write an input file in any ASCII text editor. This input file consists of keywords that define the simulation options. Most important keywords are explained in the examples of this Section, and the ones not covered here are discussed in Appendix~\ref{sec:further}. 

After creating the input file, say \verb|file.in|, one just needs to type \verb|qw2d file.in|. The results are stored in some output files. The \verb'file.dat' output file contains the final probability distribution. The \verb'file-wave.dat' output file contains the final complex amplitudes of the wave function. The \verb'file-pb.dat' output file contains the approximate stationary distribution, when it is requested. The \verb'file-screen.dat' output file contains the data observed in the screen, when it is requested. The \verb'file.sta' output file contains certain statistics such as variance, standard deviation, average and total variation distance from an approximate stationary distribution and from the uniform distribution. The \verb'file.plt' output file is a gnuplot script. The postscript files it generates depend on the options used. Typically these files are: \verb|file-3d.eps|, the 3D plot;  \verb|file-2d.eps|, the contour plot; \verb|file-screen.eps|, the observation screen pattern; \verb'file-pb.eps', the approximate stationary distribution plot. Further plots can also be manually generated by the user with some knowledge on gnuplot or any other similar tool.

\subsection{Simulation of a double-slit experiment}

\begin{figure}
\centering
 \subfigure[3D plot]{\includegraphics[width=0.54\textwidth]{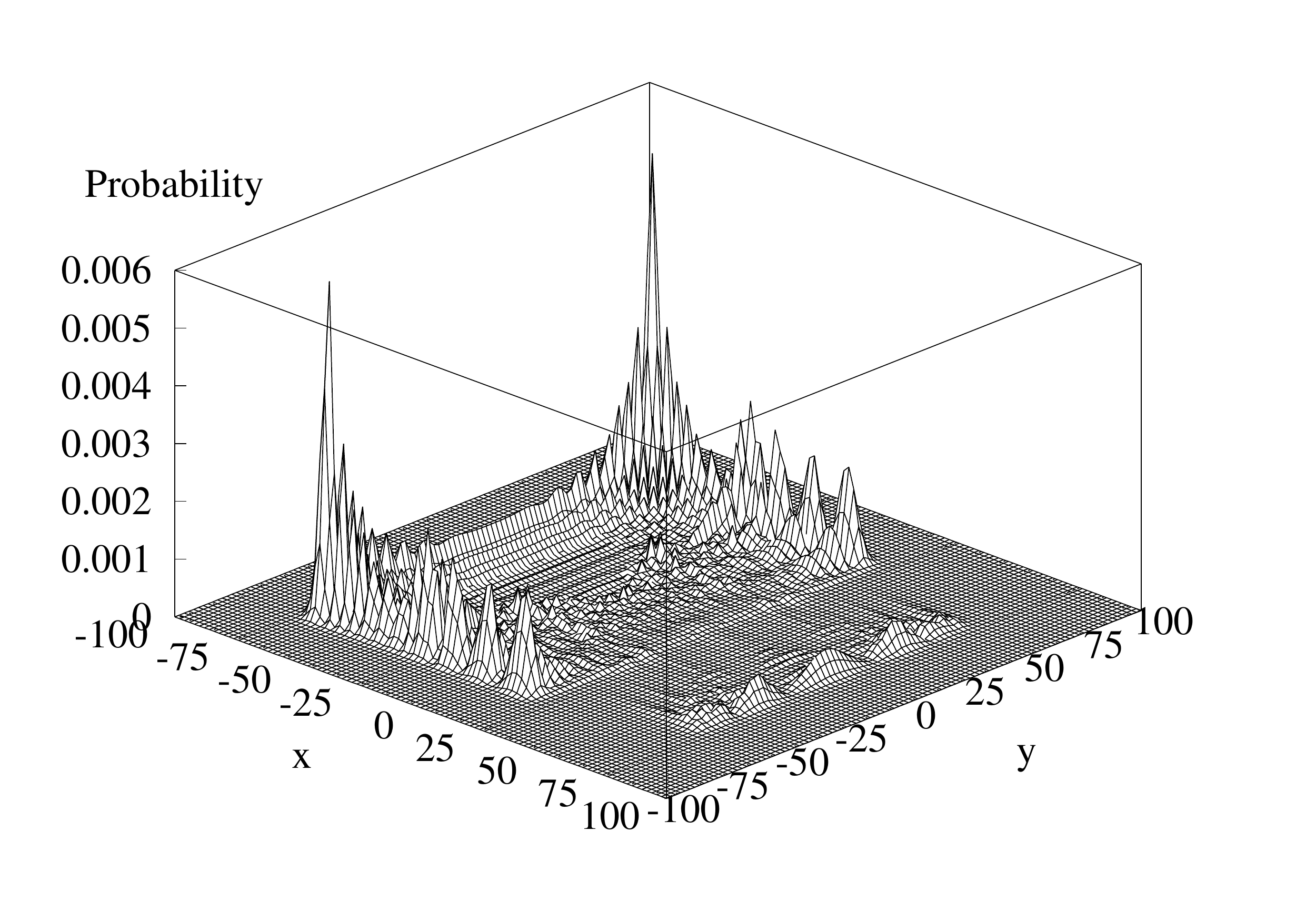}}
 \subfigure[Contour plot]{\includegraphics[width=0.45\textwidth]{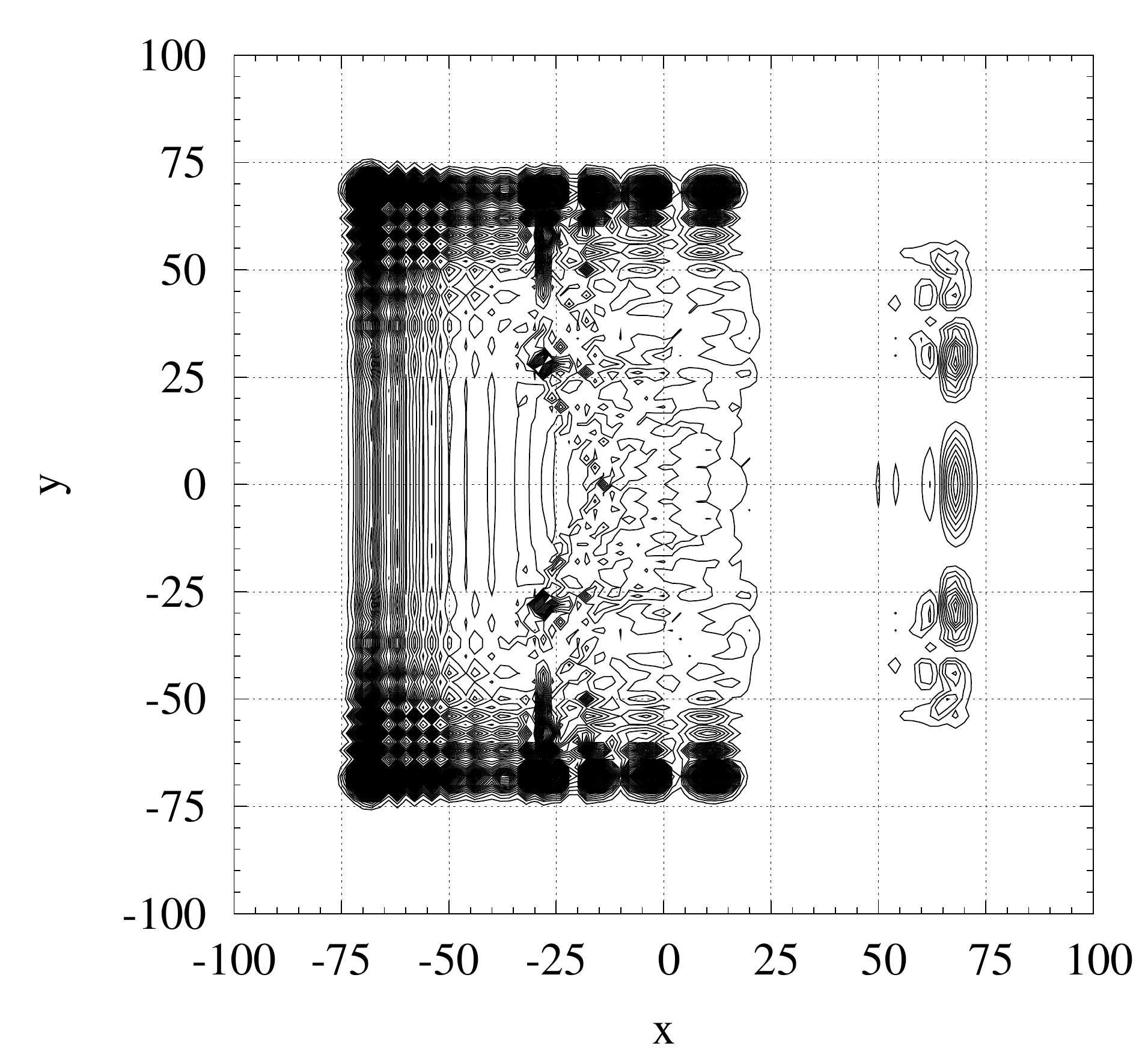}}
\caption{Probability distribution after a double-slit experiment. An amplification factor 5 was used for $x>20$, in order to improve visualization.}
\label{fig:dslit}
\end{figure}

In Fig.~\ref{fig:dslit} we have the result of a simulation of the double-slit experiment with quantum walks, reproducing some results recently obtained by Oliveira \emph{et al}~\cite{AmandaSlit07}. This simulation took less than 2s on a Pentium IV 2.6GHz with 512MB of RAM memory, 512KB of cache memory and under Linux. In order to perform this simulation with \emph{qw2d} the input file must have the following keywords, all in uppercase:
\begin{center}
%\footnotesize
\begin{minipage}{0.6\textwidth}
\begin{verbatim}
BEGIN
 COIN HADAMARD      BLPERMANENT
 STATE HADAMARD     SCREEN 60 -100 60 100
 STEPS 100          LATTYPE DIAGONAL
END
\end{verbatim}
\end{minipage}
\end{center}

It is not necessary to place the commands exactly as in the example. The user may skip lines between different commands or even write everything in a single line. It is important, however, to keep all the  keywords in the main section of the input file, delimited by a \verb|BEGIN| keyword and an \verb|END| keyword. Otherwise, the keywords will be interpreted as comments. The plots of Fig.~\ref{fig:dslit} were generated by \emph{gnuplot}, with a script provided as output of \emph{qw2d}. The generation of the plots took about 30s.

The \verb|COIN| keyword defines the coin used in the simulation. In the previous example we chose the Hadamard coin. We could also have chosen the Fourier or Grover coins, with the options \verb|FOURIER| and \verb|GROVER|, or even an arbitrary unitary coin. In the latter case, the option to be used is \verb|CUSTOM|. An extra section in the input file in also required in this case in order to define the arbitrary coin, see Appendix~\ref{sec:further}.

Analogously, the \verb|STATE| keyword  defines the initial state of the simulation. In the previous example we chose the state that provides a better spreading for the Hadamard coin. We could also have chosen the corresponding states for Grover or Fourier coins, or even an arbitrary initial state.

The \verb'STEPS' keyword defines the number of iterations that will be carried out by the simulation. In the previous example the walker performed a hundred simulation steps. The user should recall that the larger is the simulated time without boundaries the farther from the origin the particle can be found, which means that a bigger space in computer's memory must be reserved for the simulation. Therefore, this keyword may increase not only the running time but also the memory requirements of the simulation. It will be explained later on how to avoid this memory consume when we fix certain boundaries in the lattice.

The slits were created with help of \verb|BLPERMANENT| keyword in the main section of the input file. With this command we declare that some links in the simulation will be permanently broken. In order to define the position of those broken links we use a separated section in the same input file, with the commands
\begin{center}
%\footnotesize
\begin{minipage}{0.6\textwidth}
\begin{verbatim} 
BEGINBL
 LINE 20 100 20 7
 LINE 20 5   20 -5
 LINE 20 -7  20 -100
ENDBL
\end{verbatim}
\end{minipage}
\end{center}
The command \verb|LINE x0 y0 x1 y1| isolates all the points that pass through the segment from $(x0,y0)$ to $(x1,y1)$. The line of isolated points may be parallel to the axes $x$ or $y$, or may have an angle of 45 degrees with one of those. It is also possible to isolate a single point with the command \verb|POINT x0 y0|. With combinations of these two commands one can simulate not only single- and double-slit experiments but also the evolution of the walker with an enormous variety of boundaries. In this first example we have a slit at $(20,6)$ and another at $(20,-6)$.

An observation screen may be defined with the \verb'SCREEN' keyword, which must be followed by the initial and final coordinates. The screen may be defined parallel to axes $x$ or $y$, or with an angle of 45 degrees with one of those. In the example the screen goes from $(60,-100)$ to $(60,100)$.

Since the fraction of the wave that passed through the slits in the previous example was very small, the visualization obtained in the first place by the simulation turned out to be quite precarious. In order to solve that, we used the \emph{qwamplify} tool to amplify by a factor of $5$ the whole region where $x\geq 20$. This tool may be used by typing something like \verb|qwamplify file.dat [options]|. The software creates a safety copy of \verb|file.dat| and replaces it by a new one, with the appropriate part of the wave function amplified. Help concerning the available options can be obtained simply by typing \verb|qwamplify|.

\begin{figure}
\centering
 \subfigure[3D plot]{\includegraphics[width=0.54\textwidth]{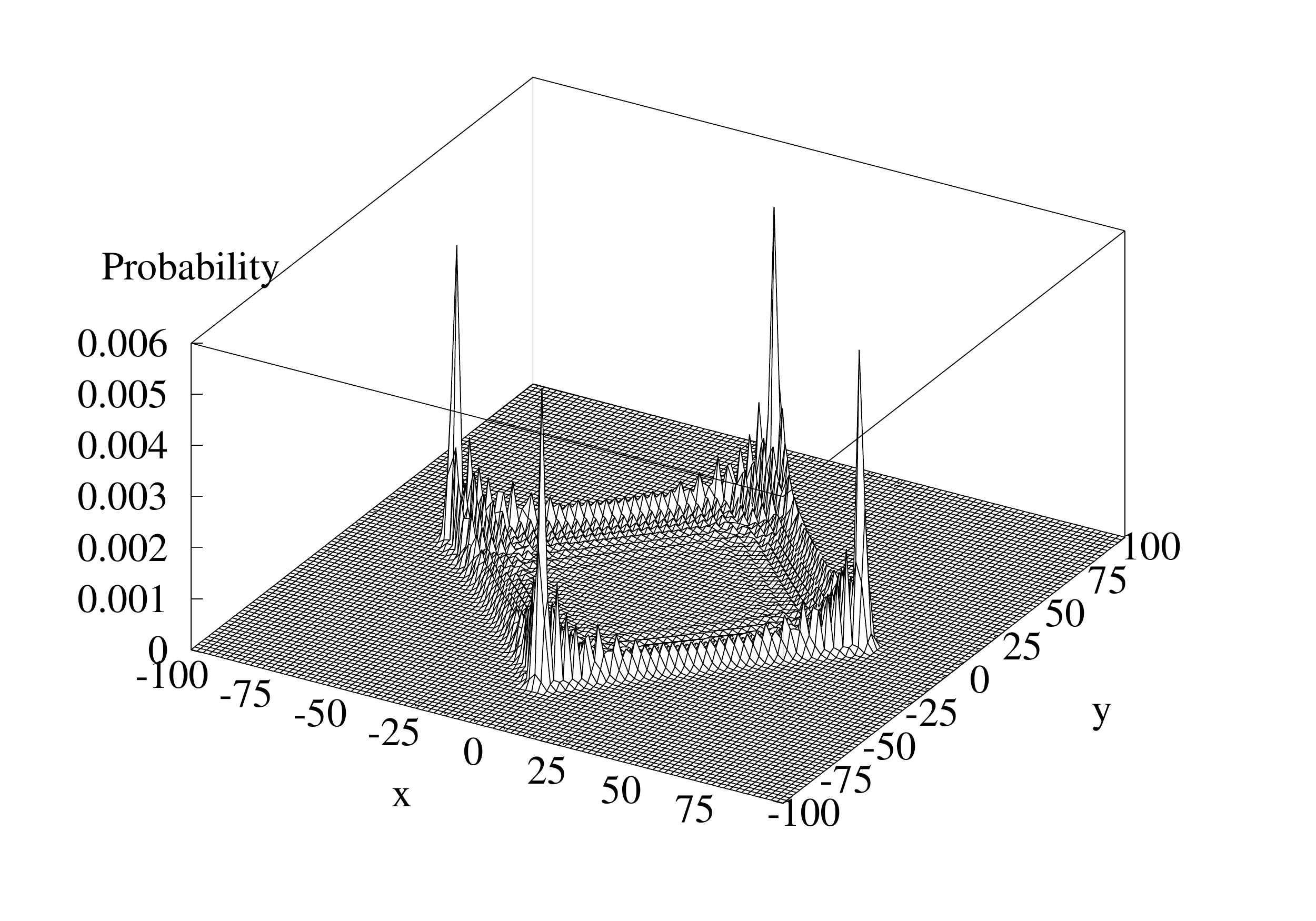}}
 \subfigure[Contour plot]{\includegraphics[width=0.45\textwidth]{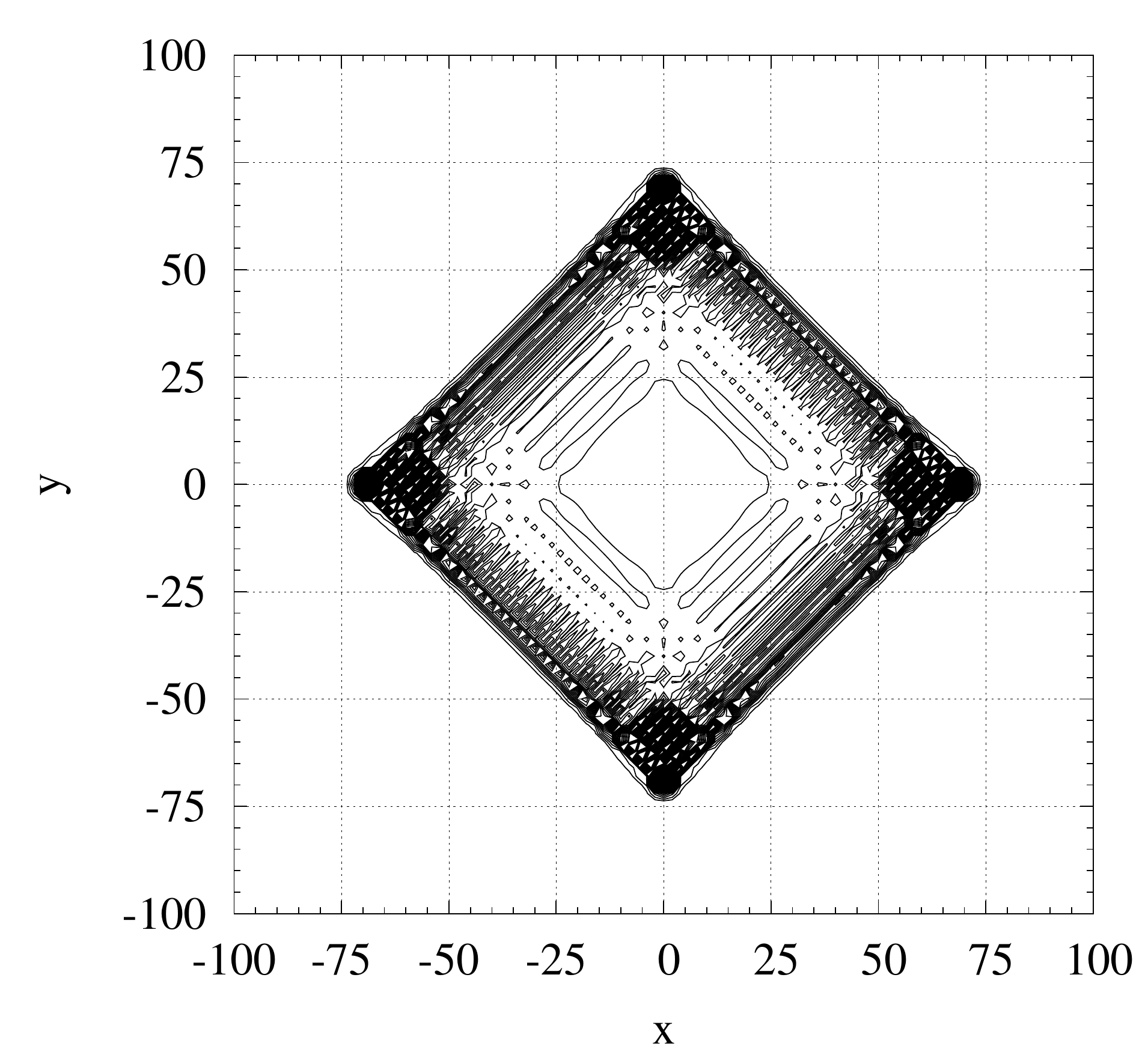}}
\caption{Probability distribution after a hundred steps of a Hadamard walker. Here, the shift operator is  such that the mathematical lattice coincides with the physical lattice.}
\label{fig:hadamardB}
\end{figure}

The \verb|LATTYPE DIAGONAL| command declares that the shift operator $S_a$ is to be used, see Eq.~\eqref{eq:shiftA}. The shift operator $S_b$, from Eq.~\eqref{eq:shiftB}, is the default but can also be explicitly declared with the \verb|LATTYPE NATURAL| command.
The natural lattice, when compared to the diagonal lattice, gives probability distributions rotated by an angle of 45 degrees. We can note this behaviour in Fig.~\ref{fig:hadamardB}, the simulation of the Hadamard walk for a hundred steps without slits. Note that there is a region on the four corners of the plot that is not used. In most situations the diagonal lattice provides better visualization than the natural lattice.
This kind of non-diagonal lattice was used by Innui \emph{et al}~\cite{KonnoIK04} with a slightly different shift operator. Our evolution equation, however, has the advantage of preserving the final probability distribution---except for the above mentioned rotation---with the same coin operator. 

\begin{figure}
\centering
\subfigure[Simulation with $T=100$ steps and screen along $x=60$.]{\includegraphics[width=0.45\textwidth]{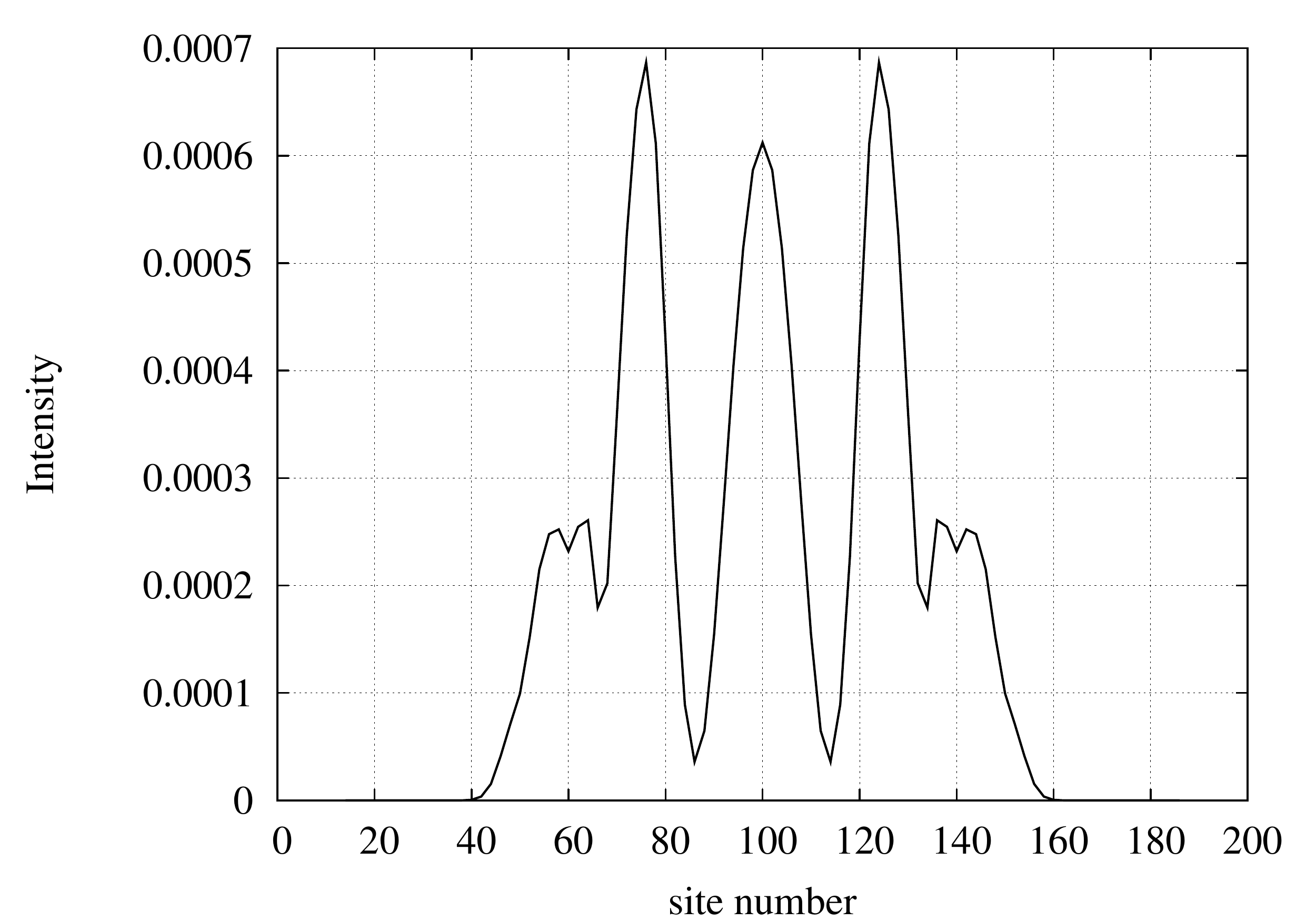}}
\subfigure[Simulation with $T=800$ steps and screen along $x=500$.]{\includegraphics[width=0.45\textwidth]{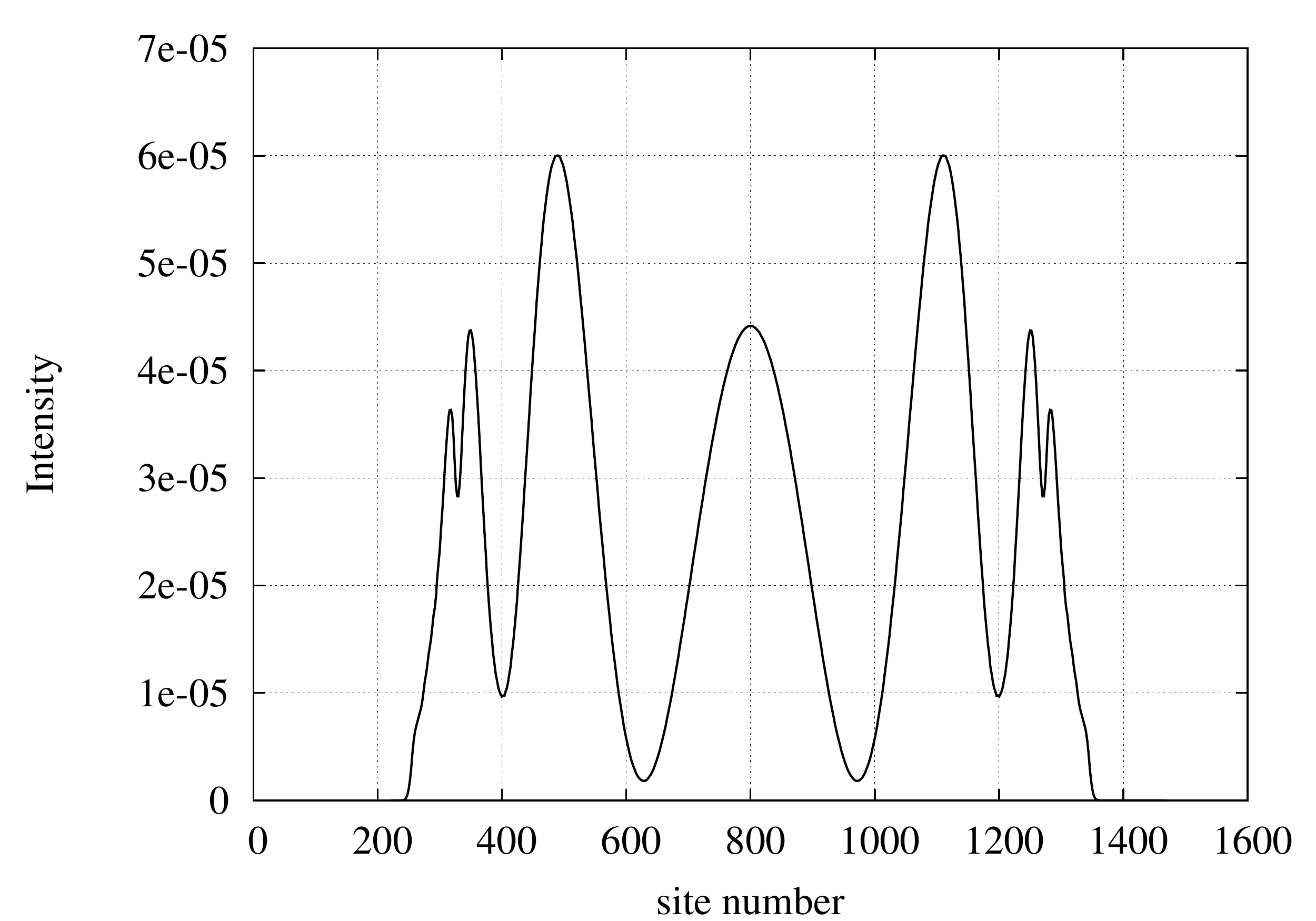}}
\caption{Simulation of observation screens in the experiment of double slit.}
\label{fig:dslit-screen}
\end{figure}

In Fig.~\ref{fig:dslit-screen} we have results of two experiments of double slit with Hadamard walkers. The slits were placed exactly as in the previous example: one at $(20,6)$ and another at $(20,-6)$. In the first simulation we carried out $T=100$ steps; in the second simulation, $T=800$ steps. The simulation took about 15min in the latter case. In the plots of Fig.~\ref{fig:dslit-screen} we have the pattern that would be observed in screens placed, respectively, along $x=60$ and $x=500$. In the latter case we note that the local minima are lower and the curve smoother.

\subsection{Simulation of detectors}

Detectors are described by a collection $\{M_m\}$ of measurement operators which satisfy the completeness equation, $\sum_m{M_m^\dagger M_m} = I$. The index $m$ indicates the result of the measurement. Before state $\ket{\Psi}$ is measured the probability that result $m$ occurs is given by $p(m)=\bra{\Psi}M_m^\dagger M_m\ket{\Psi}$, and after the measurement the state collapses to $\ket{\Psi^\prime}=\frac{1}{\sqrt{p(m)}}M_m\ket{\Psi}$.
In \emph{qw2d} we may define an arbitrary number of detectors specifying a list of coordinates $(m_1, n_1), \cdots, (m_N,n_N)$. The measurement operators are 
\begin{equation}
M_0 = I_4\otimes I_\infty - \sum_{i=1}^{N}{M_i}
\end{equation}
and 
\begin{equation}
M_i = I_4 \otimes \ket{m_i, n_i}\bra{m_i, n_i},
\end{equation}
for $1\leq i \leq N$.

\begin{figure}
\centering
\subfigure[Contour plot of the final probability distribution]{\includegraphics[width=.43\textwidth]{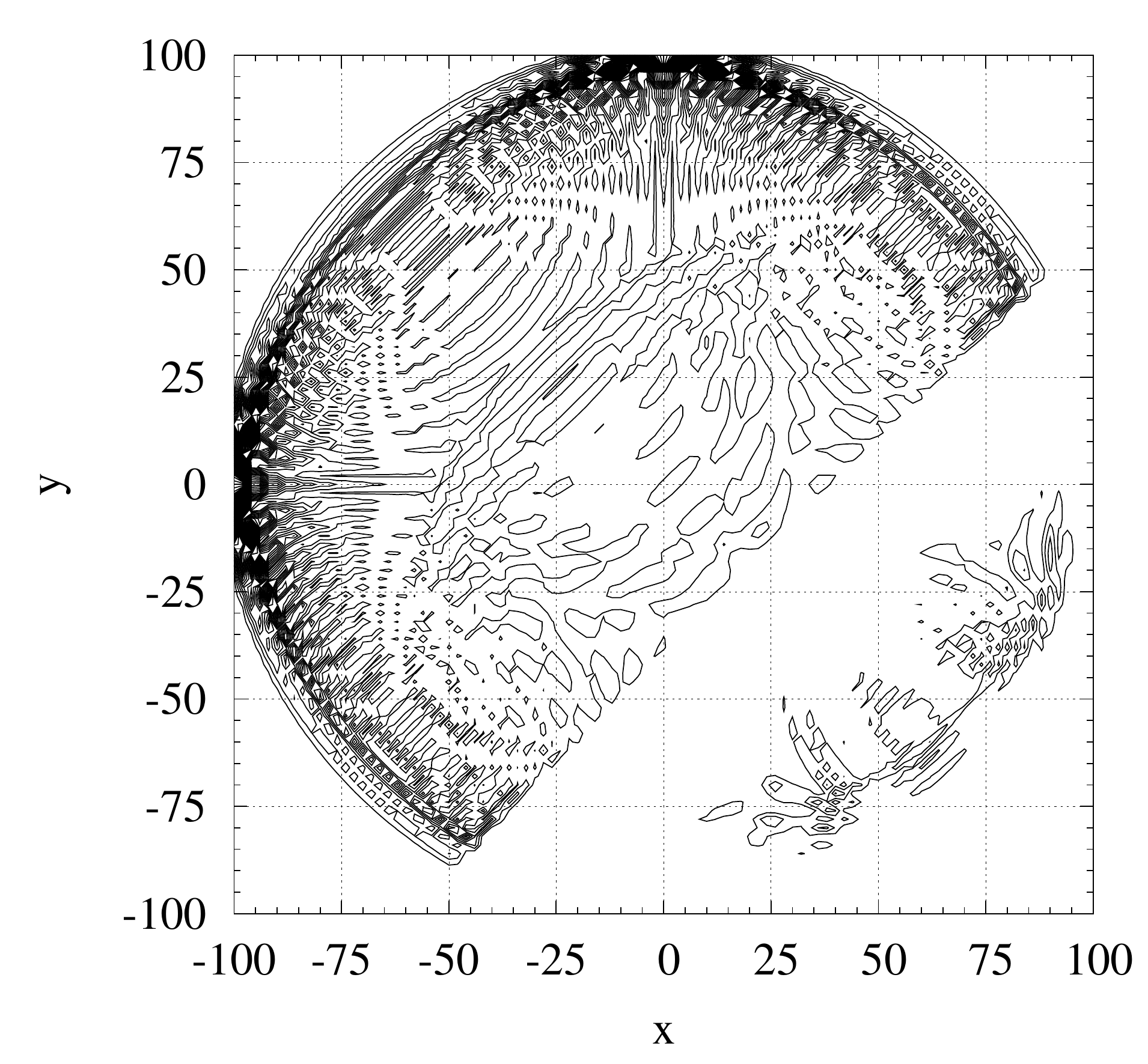}}
\subfigure[Simulation of observation screen]{\includegraphics[width=.55\textwidth]{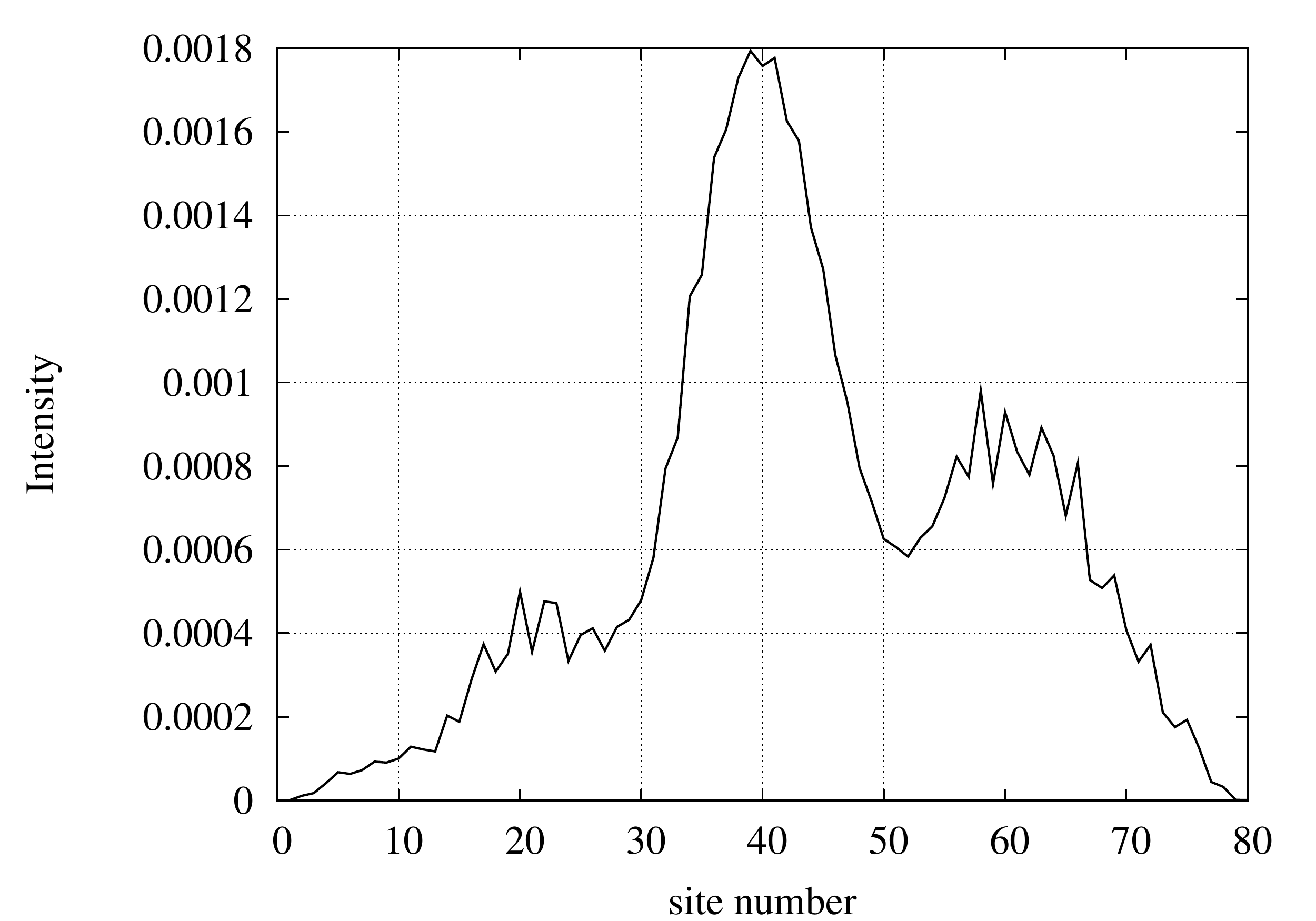}\label{fig:dslit-diag-screen}}
\caption{Results of a double-slit experiment with Grover walker. Both the wall and the observation screen are parallel to the secondary diagonal, and there is a detector near one of the slits.}
\label{fig:dslit-diag}
\end{figure}

In Fig.~\ref{fig:dslit-diag} we have the plots of another double-slit experiment. In this simulation we use the Grover coin and place the wall parallel to the secondary diagonal. The wall goes from $(-60,-100)$ to $(100,60)$. One slit goes from $(13,-27)$ to $(15,-25)$ and the other goes from $(25,-15)$ to $(27,-13)$. We have also placed a detector near the first slit, at $(15,-27)$. The experiment was repeated ten times in order to take the average of the results. 

The positions of the detectors are defined in the main section of the input file, by using the \verb|DETECTORS| keyword followed by the number of detectors and their respective coordinates. The number of repetitions of the experiment is defined by using the \verb|EXPERIMENTS| keyword. Here we have an example on how to use these two keywords:
\begin{center}
%\footnotesize
\begin{minipage}{0.6\textwidth}
\begin{verbatim} 
DETECTORS 1 15 -27   
EXPERIMENTS 10
\end{verbatim}
\end{minipage}
\end{center}

The observation screen, also placed parallel to the secondary diagonal, goes from $(20,-100)$ to $(100,-20)$. The $x$ axis in Fig.~\ref{fig:dslit-diag-screen} is labeled sequentially, starting from the first point of the screen. We note in both plots that the interference pattern was asymmetric and also weaker on the detector's side.

\subsection{Simulation of finite lattices}

The QWalk simulator can be used to simulate quantum walks on finite lattices. In this Section we study the example of a square lattice but QWalk also allows the definition of arbitrary boundaries. In the case addressed here, the boundaries were generated by breaking the links over a square the vertices of which are $(-M,M)$, $(M,M)$, $(M,-M)$ and $(-M,-M)$, as in~\cite{AmandaWeciq06}. Different values of $M$ were investigated. The stationary distribution was approximated by running the simulation for 5000 steps.  

In order to calculate the total variation distance we must have the \verb|MIXTIME| keyword in the main section of the input file, followed by the number of steps used to approximate the stationary distribution. This number should be higher than---or at least equal to---the number of steps that will be simulated, and the positions of permanent broken links must be correctly declared in order to define a closed region of the lattice. Since the number of steps simulated may be much greater than the size of the achievable lattice, we may improve the performance of the simulation by using the \verb|LATTSIZE| keyword. This keyword declares that the lattice is only used from from ${x=-max}$ to ${x=max}$ and from ${y=-max}$ to ${y=max}$, where $max$ is the integer number passed as argument of the \verb|LATTSIZE| keyword. This option saves both memory and running time and should be used whenever the walk is restricted to a finite region of the lattice. The \verb'LATTSIZE' keyword must be used \emph{after} the \verb'STEPS' keyword.

For the previous example, taking $M=60$, we may prepare an input file with the keywords
\begin{center}
%\footnotesize
\begin{minipage}{0.6\textwidth}
\begin{verbatim} 
MIXTIME 5000
STEPS 2000
LATTSIZE 59
\end{verbatim}
\end{minipage}
\end{center}
together with the keywords declaring the boundary,
\begin{center}
%\footnotesize
\begin{minipage}{0.6\textwidth}
\begin{verbatim} 
BEGINBL
 LINE -60 60 60 60    LINE 60 60 60 -60
 LINE 60 -60 -60 -60  LINE -60 -60 -60 60
ENDBL
\end{verbatim}
\end{minipage}
\end{center}

When the \verb|MIXTIME| option is used, the approximate stationary distribution $\pi_{app}$ is obtained in the beginning of the simulation. Afterwards the \emph{.sta} output file keeps the total variation distance, at each step, from the average distribution to both the stationary and the uniform\footnote{In its present form, the simulator calculates the uniform distribution over all sites of the mathematical lattice, not taking into consideration the particular boundary.} distributions, so that the user may easily plot this information with an appropriate software such as gnuplot.

\begin{figure}
\centering
\subfigure[Total variation distance]{\includegraphics[width=.45\textwidth]{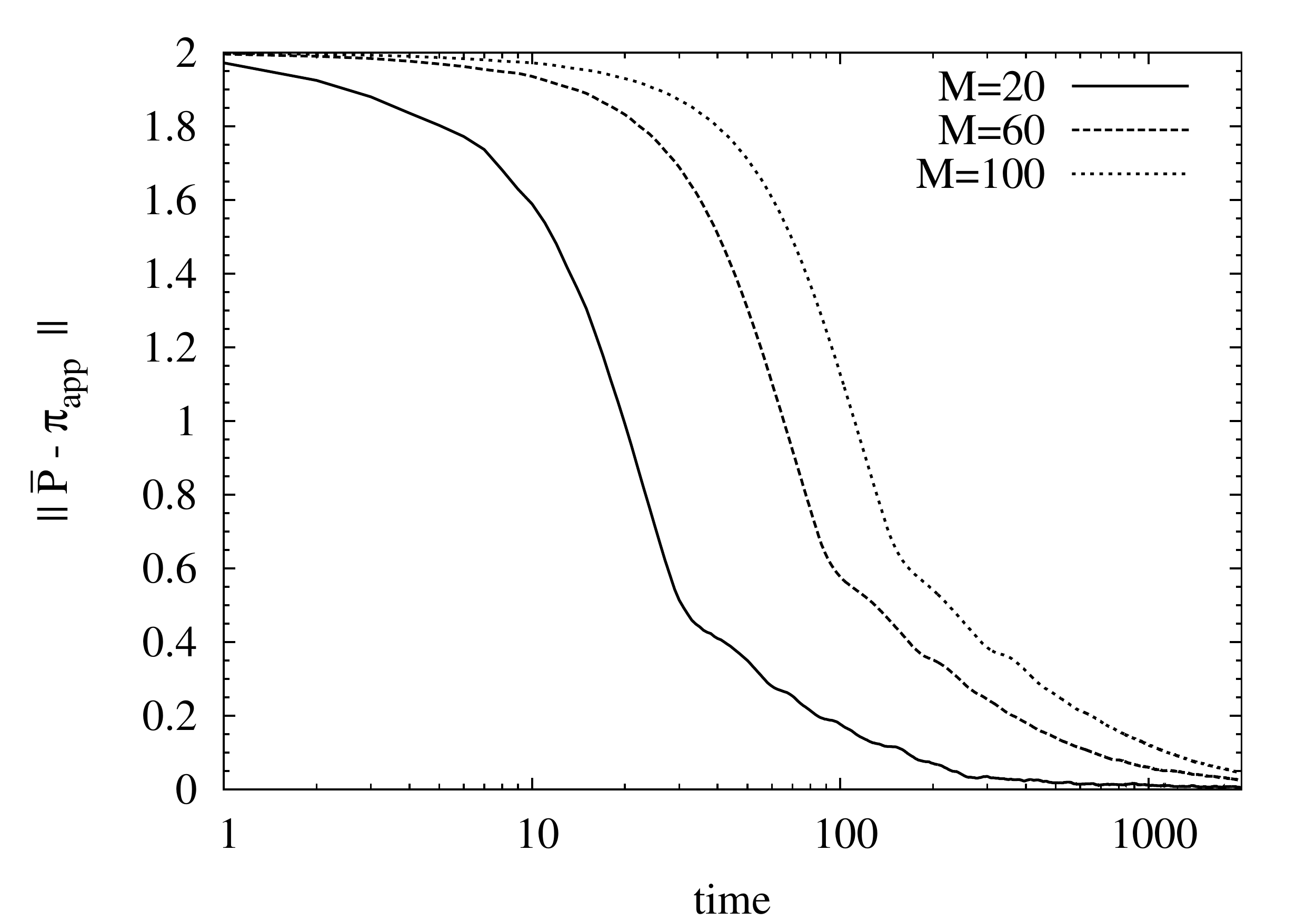}\label{fig:had-tvd}}
\subfigure[Standard deviation]{\includegraphics[width=.45\textwidth]{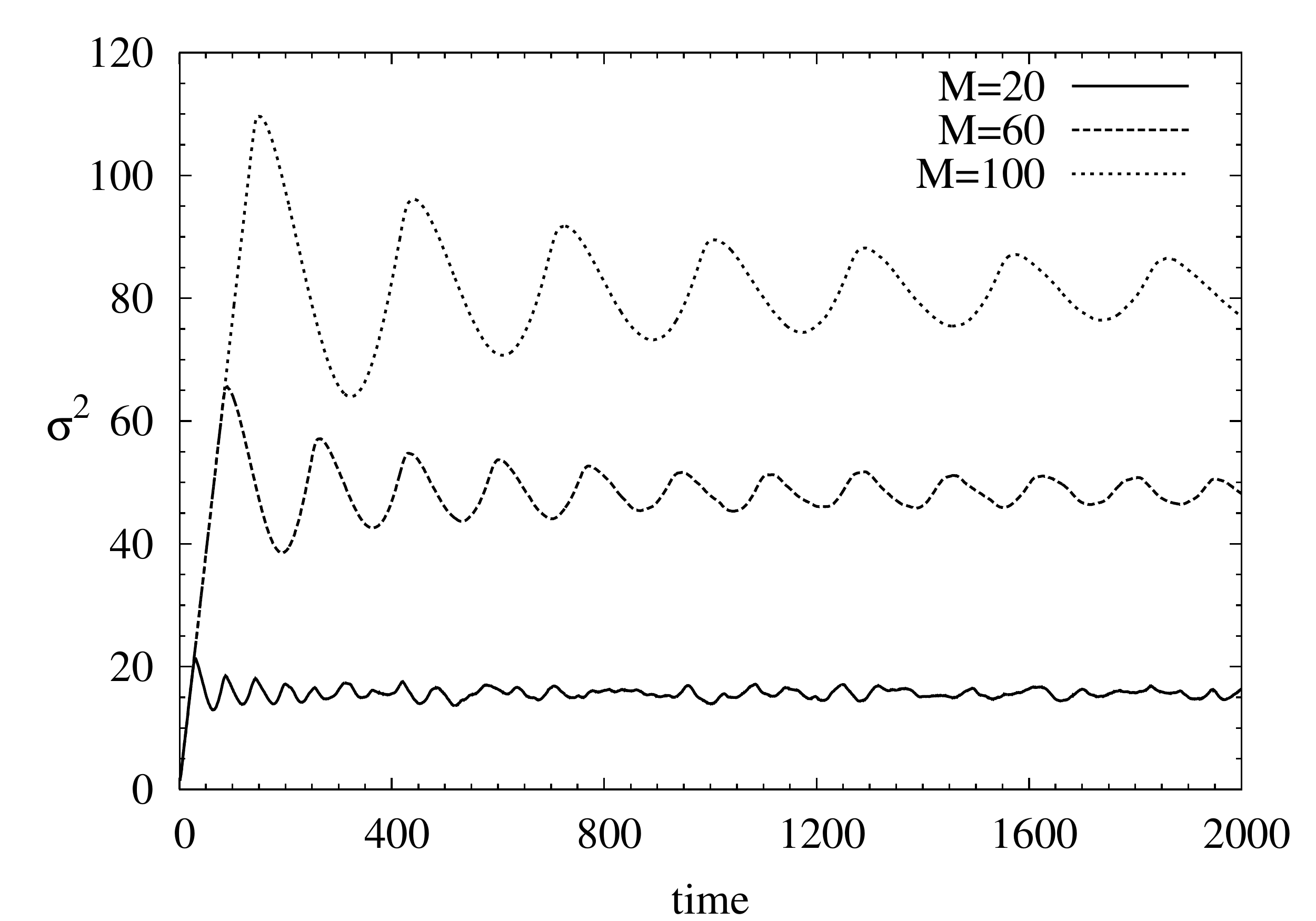}\label{fig:had-std}}
\caption{Total variation distance between the average distribution and the approximate stationary distribution as a function of time; and evolution of the standard deviation. Both plots refer to the Hadamard walk in the diagonal lattice and for different sizes of square boxes.}
\label{fig:had-tvdstd}
\end{figure}

In Fig.~\ref{fig:had-tvd} we have the total variation distance between the average distribution and the approximate stationary distribution as a function of time, for the Hadamard walk in a finite lattice with rectangular boundaries, for different values of $M$. 
In Fig.~\ref{fig:had-std} we have the evolution of the standard deviation of the Hadamard walk in the diagonal lattice for different values of $M$. We find in the \emph{.sta} output file enough information to generate this kind of plot with just a few gnuplot commands, or with any other similar tool. We should note that these plots are also consistent with the results obtained by Oliveira \emph{et al}~\cite{AmandaWeciq06}.

\begin{figure}
\centering
\includegraphics[width=.6\textwidth]{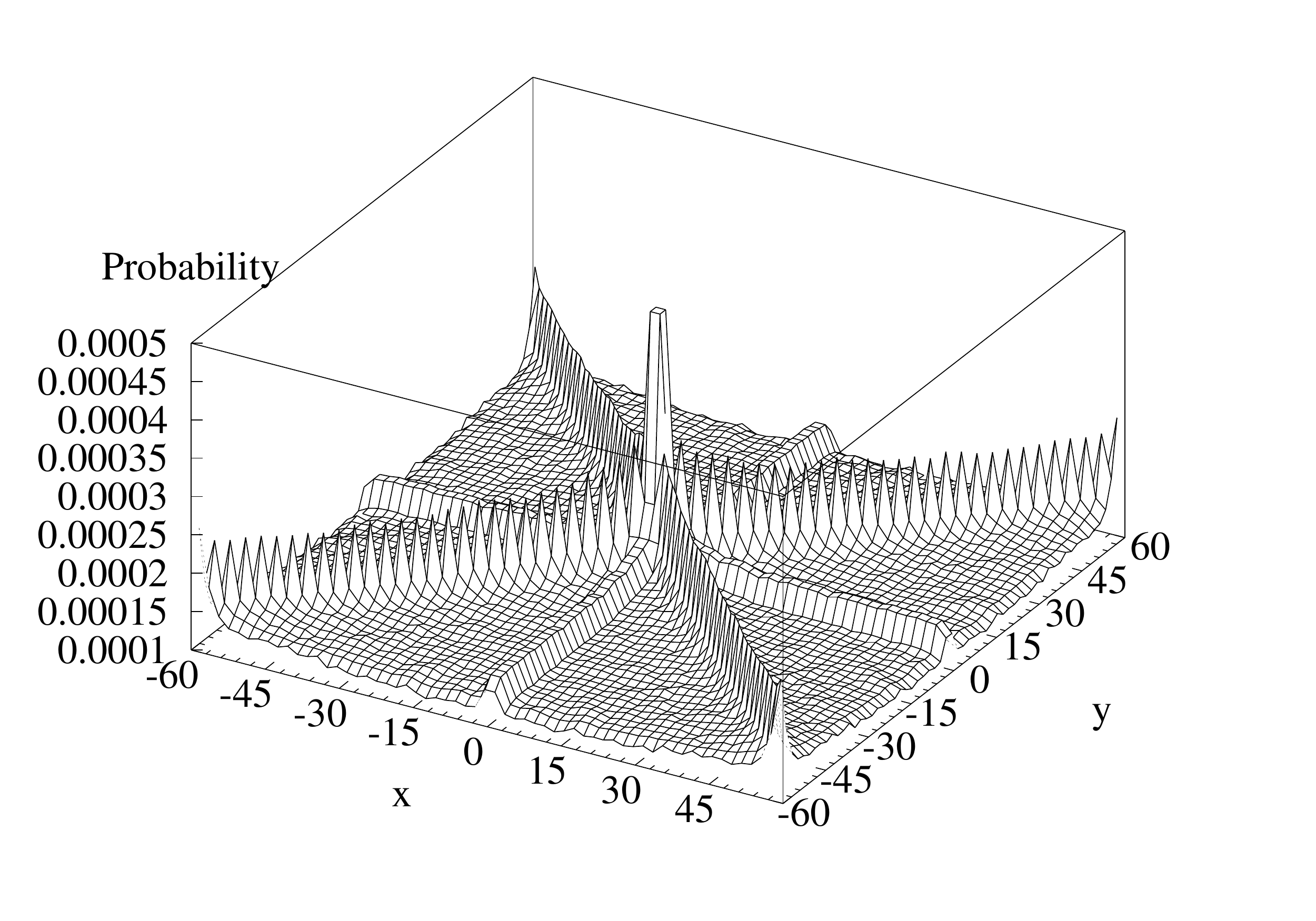}
\caption{Stationary distribution approximated with 5000~steps and a box with $M=60$.}
\label{fig:box-stat}
\end{figure}

In Fig.~\ref{fig:box-stat} we have the approximate stationary distribution for the Hadamard walk inside a box with $M=60$, which is visually different from the uniform distribution. Moreover, one may also confirm with the output file from QWalk that the average distribution does not converge to the uniform distribution even for large times.

\subsection{Simulation of decoherence}

QWalk allows the simulation of two-dimensional quantum walks with two different sources of decoherence. The first of them is generated by measurements from randomly disposed detectors. It is also possible to simulate a unitary noise~\cite{Shapira+03} which randomly opens links of the lattice. This noise model has been studied for the quantum walk on a line~\cite{Abal+05} and on a plane~\cite{AmandaPhysRev06}.

\begin{figure}
\centering
\includegraphics[width=.5\textwidth]{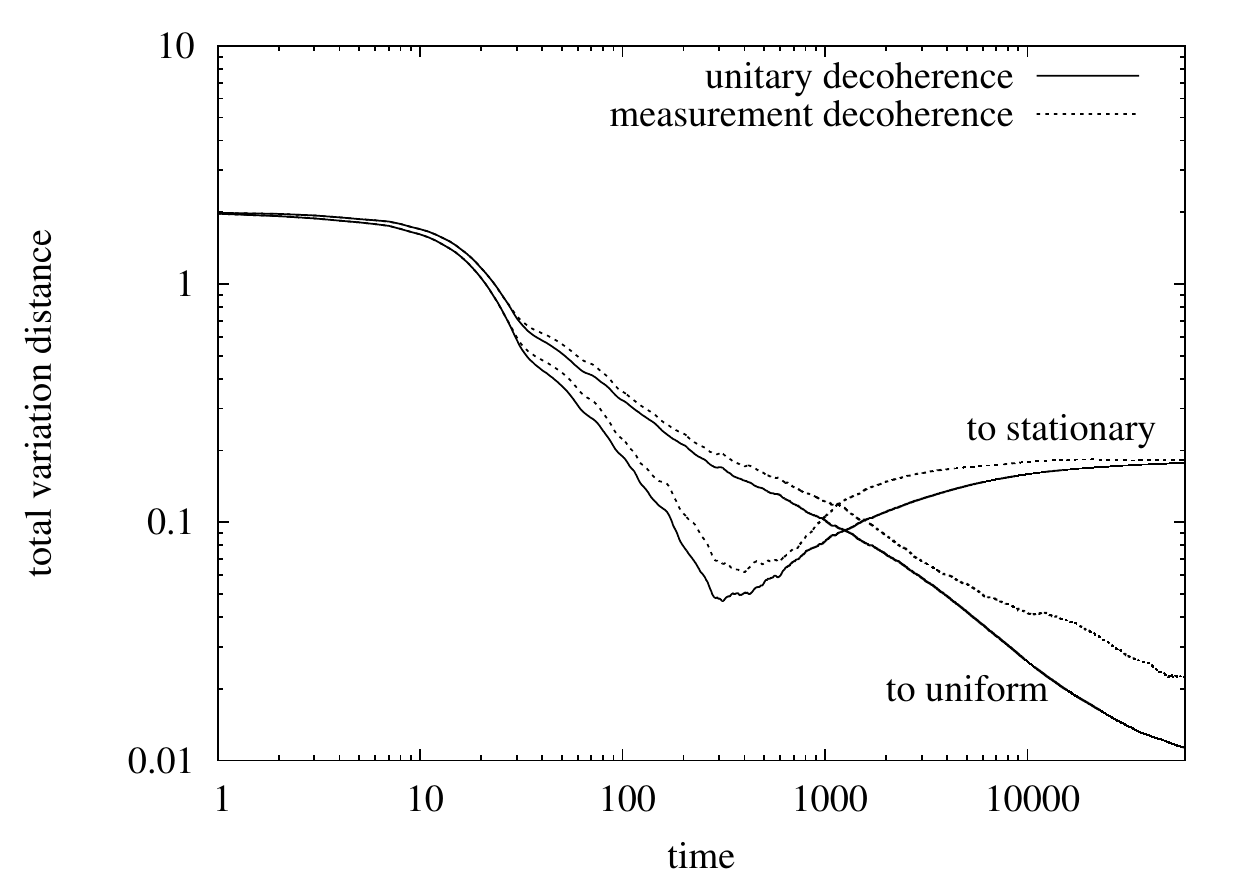}
\caption{Total variation distance as a function of time from the average distribution for the decoherent two-dimensional walk to both the uniform and the coherent stationary distributions. Two sources of noise are compared. The Hadamard coin was used and the stationary distribution was approximated with $7\cdot 10^4$ steps.}
\label{fig:decohere-compare}
\end{figure}

The measurement decoherence is better observed in the simulation of finite lattices. In order to declare the probability of measuring each site we use \verb'DTPROB' keyword in the main section of the input file, followed by a numerical value. In Fig.~\ref{fig:decohere-compare} we compare two kinds of noises---measurements and broken links---with the same decoherence parameter $p=0.01$ and the same box size $M=20$. In both cases the experiment was repeated ten times with the Hadamard coin in order to take the average of the results. In the dotted lines we have the results for the measurement decoherence and in the full lines we have the results for the unitary decoherence. In both cases we have the total variation distance, as a function of time, between the average distribution and both the uniform and the coherent stationary distributions. The stationary distribution was approximated with $7\cdot 10^4$ steps. We note that in both cases the average distribution initially approaches the coherent stationary distribution until $\sim 1/p$~time steps, going to the uniform distribution after that.

\begin{figure}
\centering
\subfigure[3D plot]{\includegraphics[width=.5\textwidth]{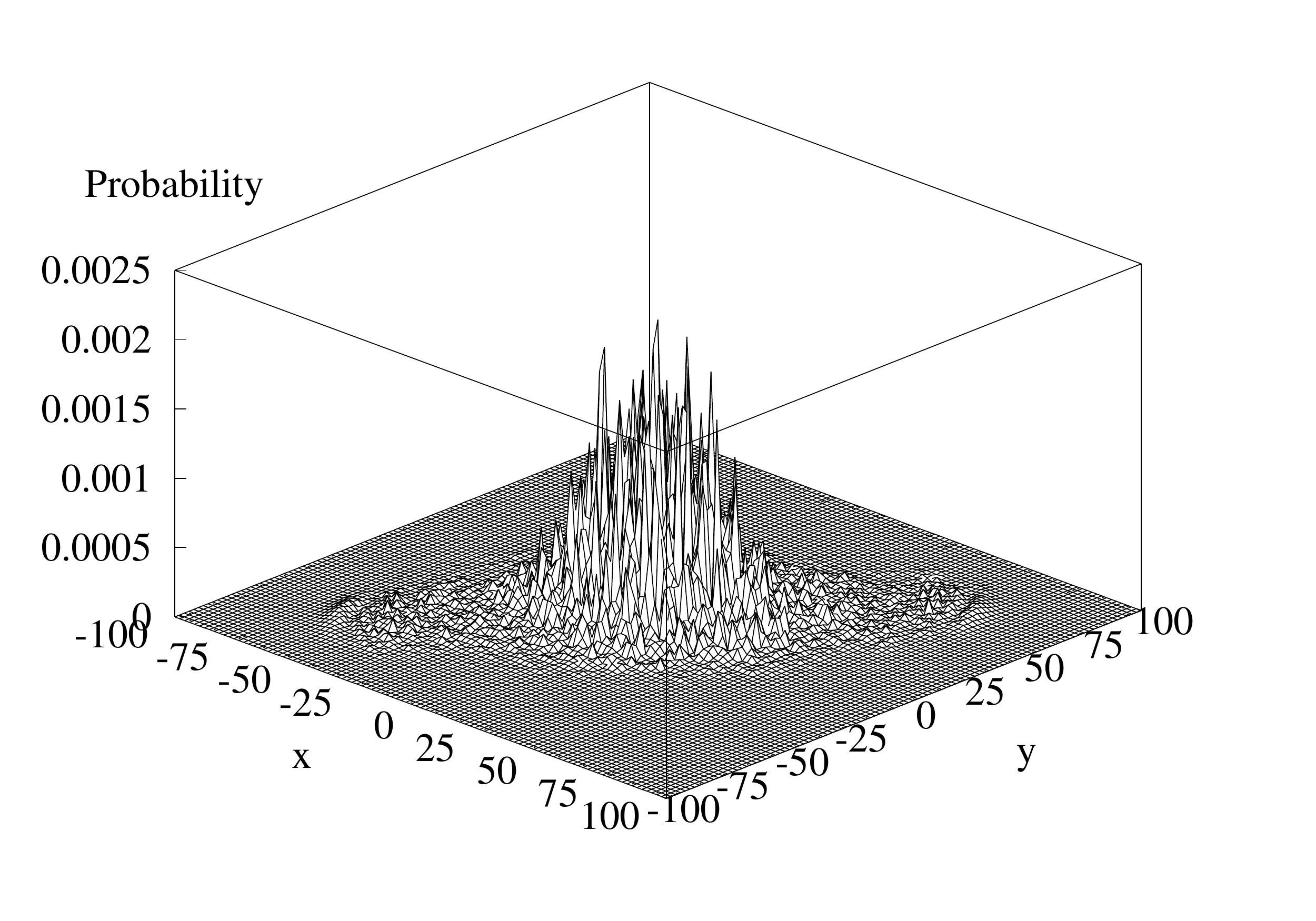}}
\subfigure[Contour plot]{\includegraphics[width=.45\textwidth]{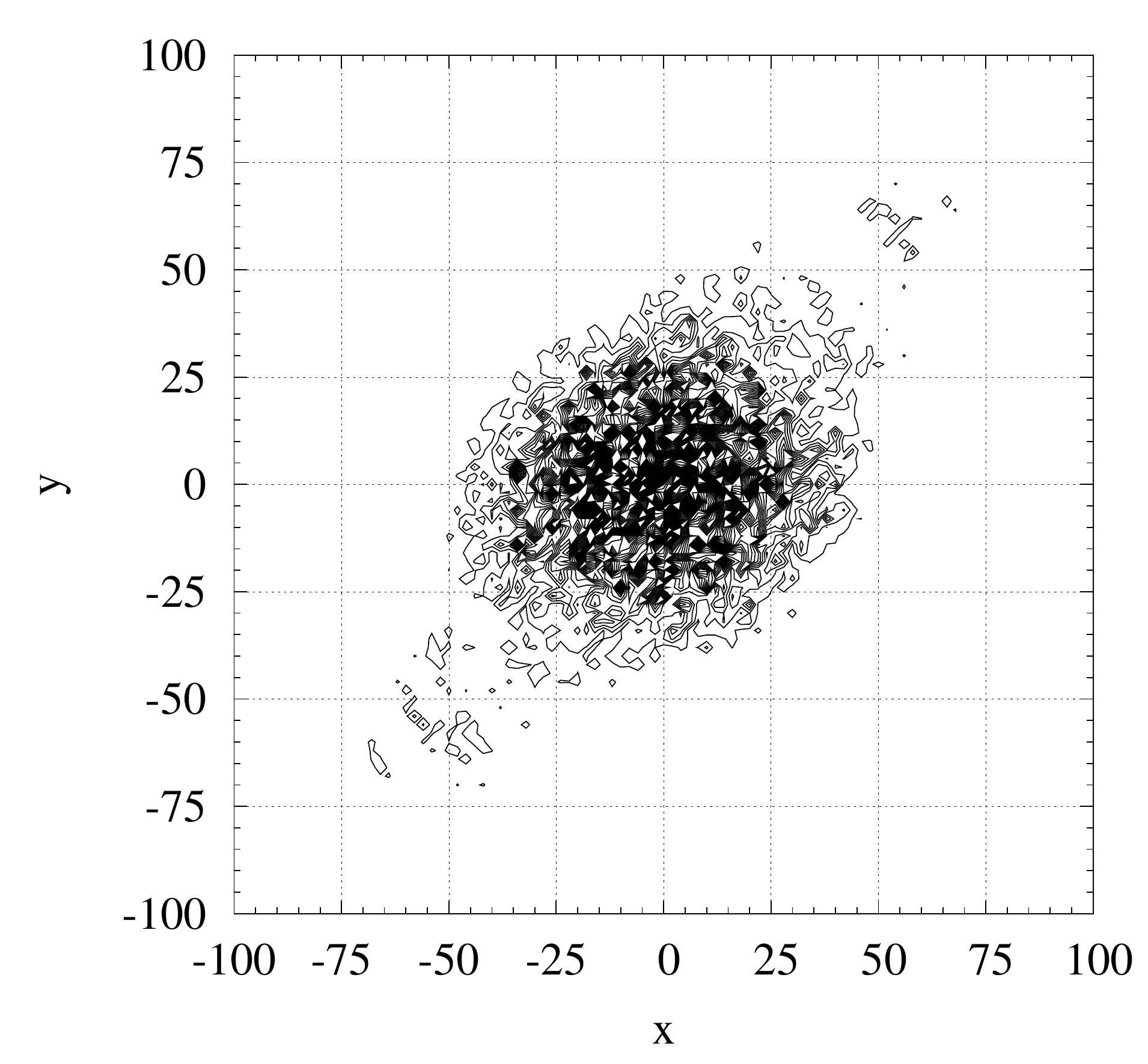}}
\caption{Probability distribution after a hundred steps of a decoherent Fourier walker in the diagonal lattice. The probability of broken links was asymmetric, namely $p_0=0$ in the secondary diagonal and $p_1=0.2$ in the main diagonal.}
\label{fig:decoherent2d}
\end{figure}

In Fig.~\ref{fig:decoherent2d} we have the result of a simulation of a Fourier walk with asymmetric unitary decoherence. The results are similar to those obtained by~\cite{AmandaPhysRev06}. We declared a probability $p_0=0$ of broken links in the secondary diagonal, and a probability of $p_1=0.2$ of broken links in the main diagonal. This kind of simulation can be done with the \verb|BLPROB| keyword, which must be followed by the probabilities of broken links in each direction---in the secondary and the in main diagonals, when the diagonal lattice is used; in the horizontal and in the vertical directions, when the natural lattice is used instead.

\subsection{Simulations in one-dimensional lattices}

The usage of \emph{qw1d} is analogous to that of \emph{qw2d}, except for the absence of some keywords. Namely, \emph{qw1d} does not recognize \verb|SCREEN|,  \verb|BLPERMANENT| and  \verb|DETECTORS|  keywords and does not recognize \verb|FOURIER| and \verb|GROVER| sub-options for coins and states. It does not recognize \verb'DIAGONAL' or \verb'NATURAL' sub-options for \verb'LATTYPE' either. On the other hand \emph{qw1d} works with three kinds of lattices, selected with \verb'LATTYPE' keyword: the infinite line is selected by \verb'LINE' sub-option; the finite one-dimensional lattice with reflecting boundaries is selected by \verb'SEGMENT' sub-option; and the cycle is selected by \verb'CYCLE' sub-option. Further information on the usage of \emph{qw1d} is provided together with the source (or binary) files.

\begin{figure}
\centering
\subfigure[Simulation with $T=1000$ steps and $p=0$]{\includegraphics[width=0.46\textwidth]{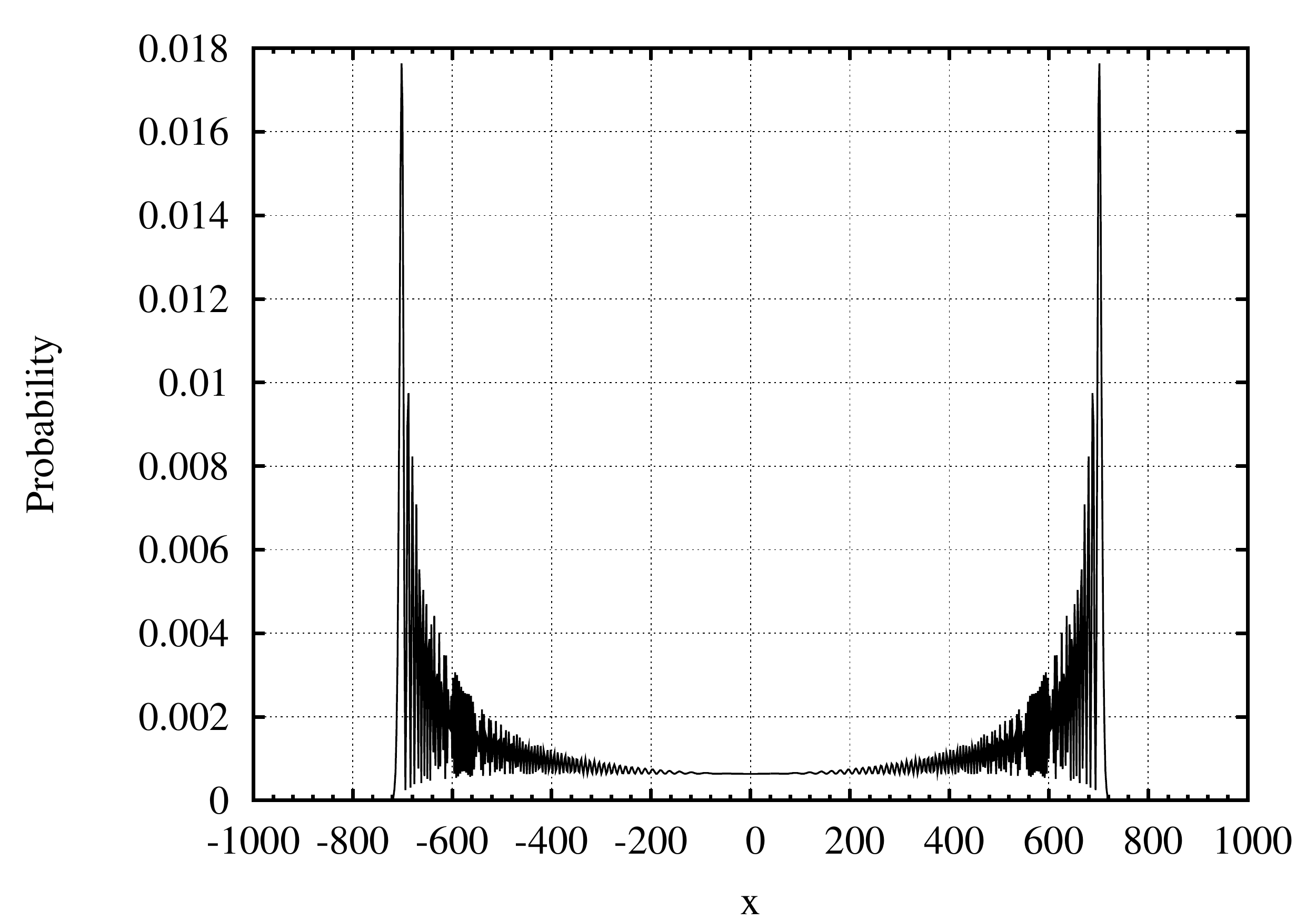}}
\subfigure[Simulation with $T=1000$ steps and $p=0.01$, taking an ensemble of $100$ experiments]{\includegraphics[width=0.46\textwidth]{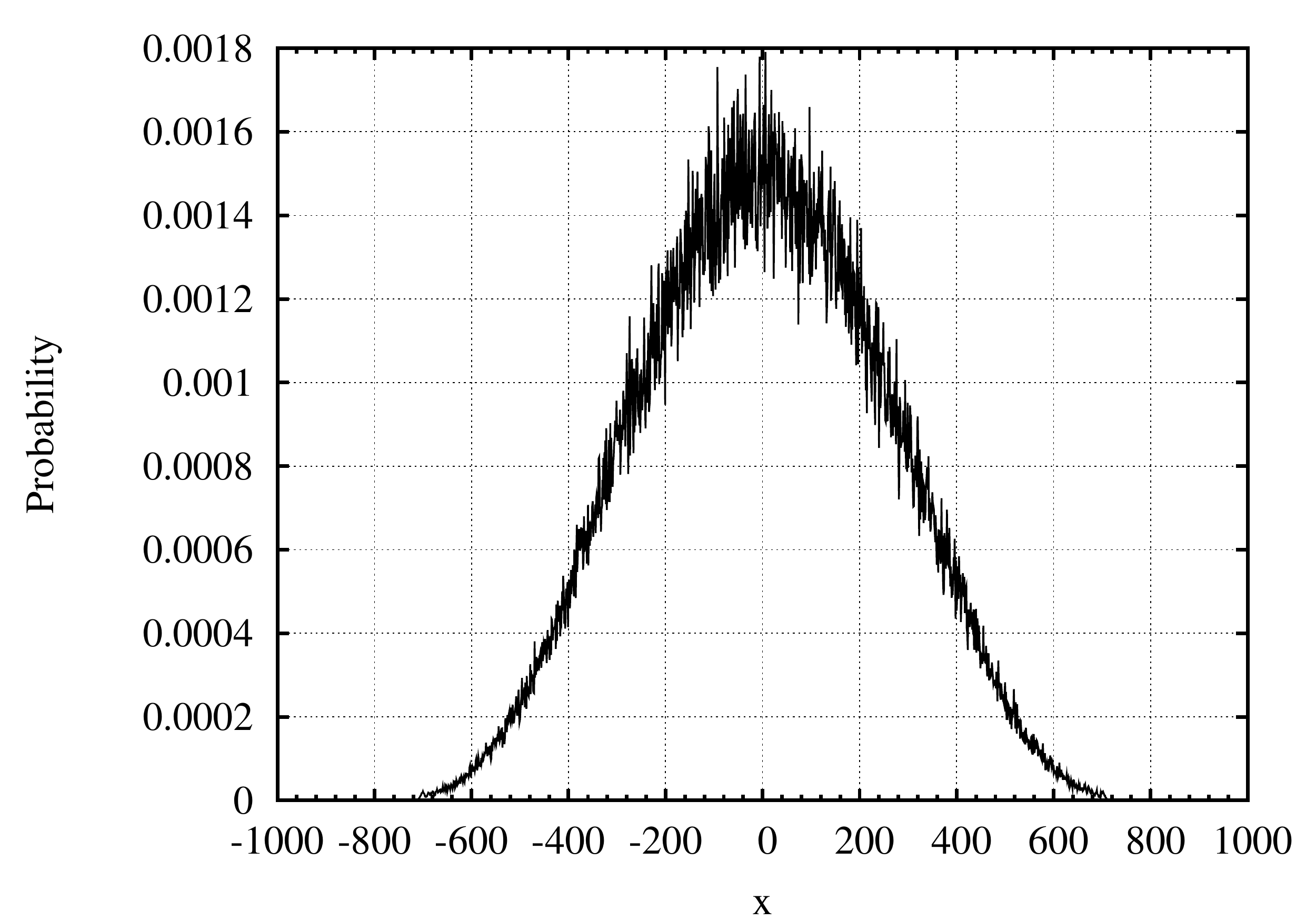}}
\caption{Quantum walk on a one-dimensional lattice with broken links.}
\label{fig:1D}
\end{figure}

In Fig.~\ref{fig:1D} we have the probability distribution for a Hadamard walk in a one-dimensional infinite lattice~\cite{KendonTregenna03,Ambainis+01,Konno02,NayakV00}. We compare the coherent-evolution case with a decoherent one, where decoherence is introduced by random broken links.
We could have introduced decoherence by means of random measurements as well. The probability of broken links in the example is $p=0.01$. In both the coherent and in the decoherent cases we have executed $T=1000$ steps with \emph{qw1d}. In the latter case we have taken as result the average over a hundred independent experiments. We note in the figure that, due to the number of steps being much higher than $1/p$, the classical behaviour of the quantum walker has already emerged~\cite{Abal+05}.

\begin{figure}
\centering
\subfigure[Final probability distribution after $T=2\cdot 10^{4}$ steps]{\includegraphics[width=0.46\textwidth]{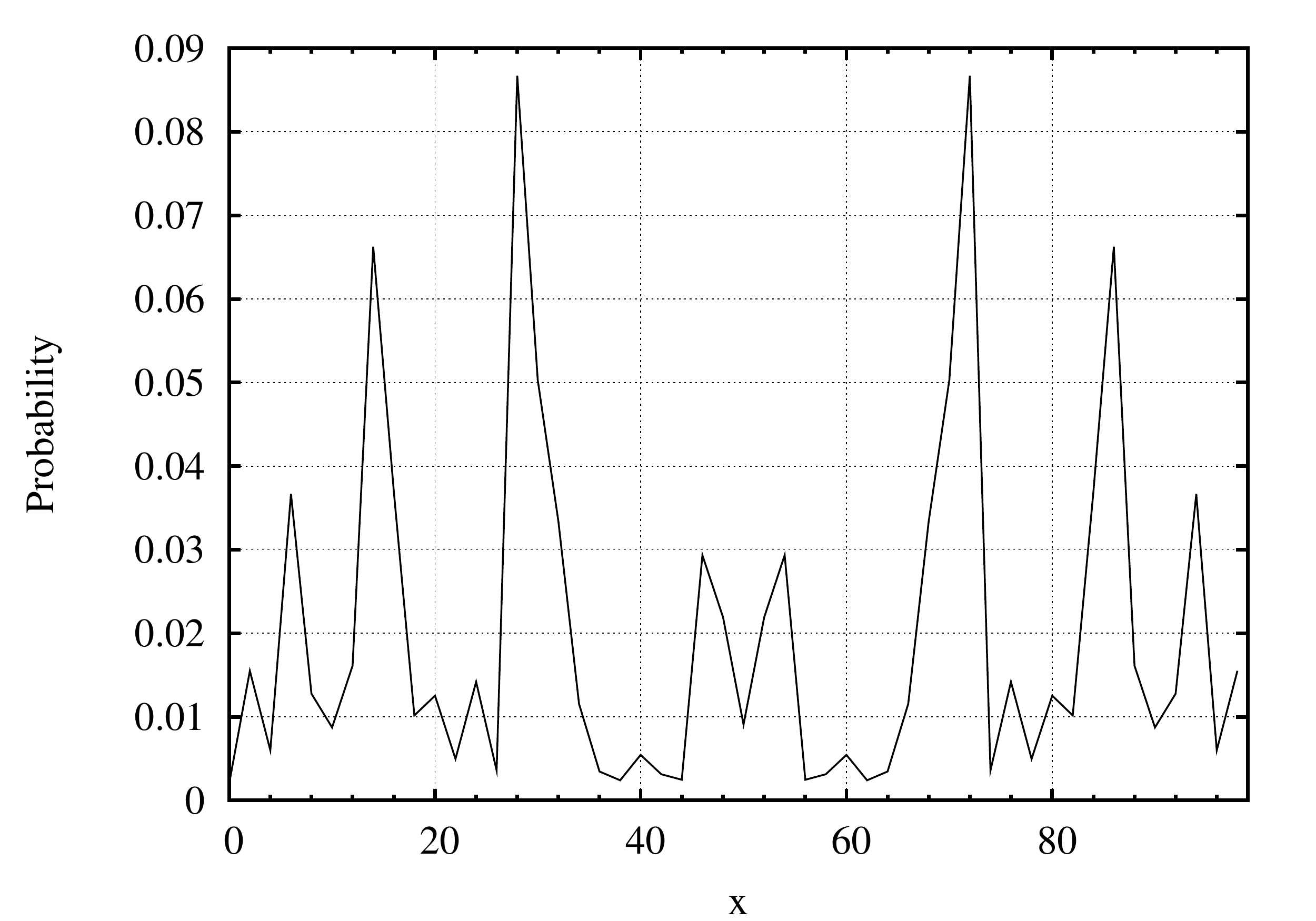}\label{fig:cycle-final}}
\subfigure[Stationary distribution approximated with $T=10^{5}$ steps]{\includegraphics[width=0.46\textwidth]{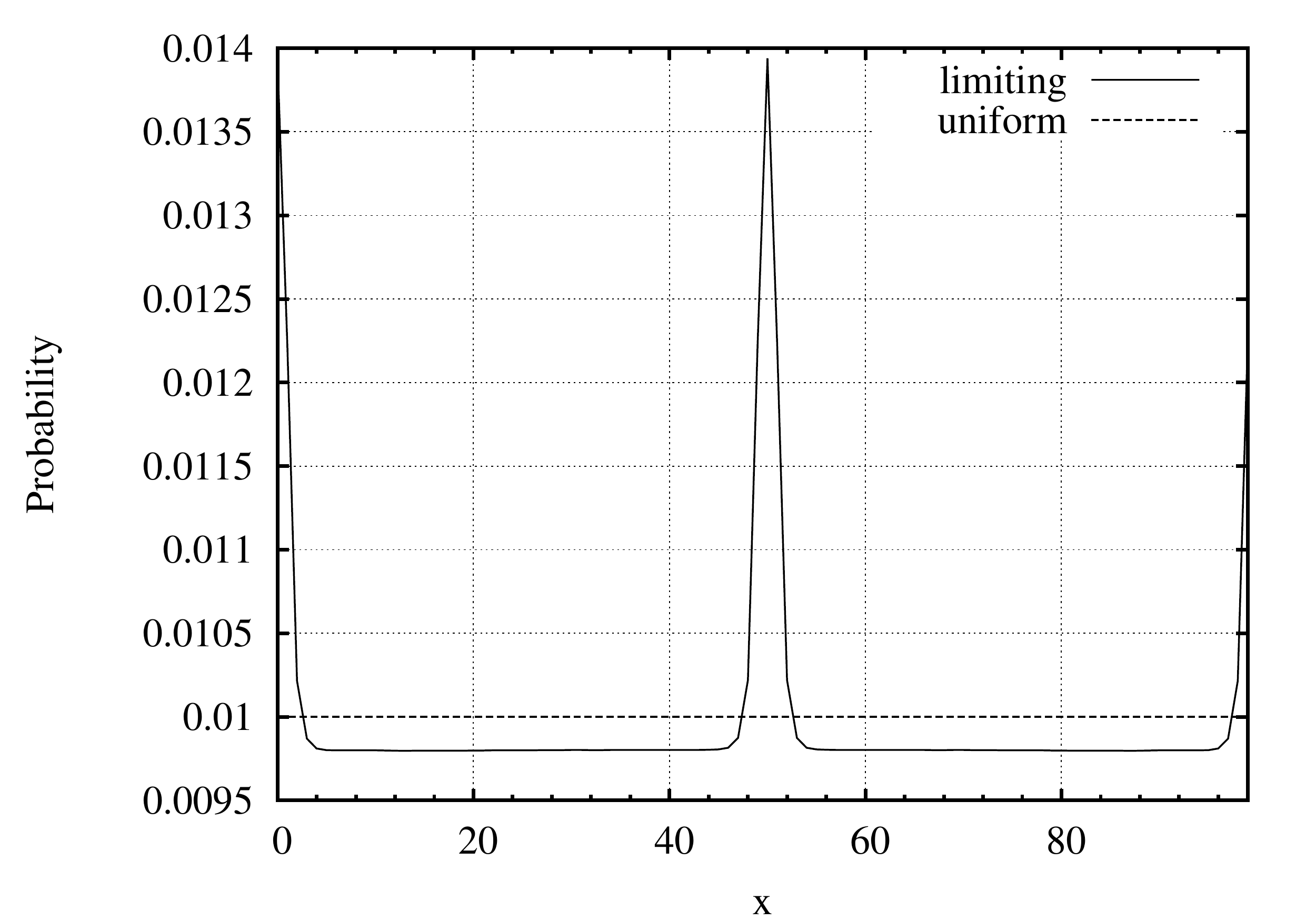}\label{fig:cycle-stat}}
\caption{Quantum walk on the cycle with a hundred sites.}
\label{fig:cycle}
\end{figure}

In Fig.~\ref{fig:cycle} we have the result of a Hadamard walk on the cycle~\cite{Kendon06,KendonTregenna03} with a hundred sites after $T=2\cdot 10^{4}$ time steps. Aharonov \emph{et al}~\cite{Aharonov+01} proved that the quantum walk on a cycle with an odd number of sites mix to the uniform distribution. The same, however, does not hold in general for cycles with an even number of sites. This is clear in Fig.~\ref{fig:cycle-stat}, where the stationary distribution was approximated with a large number of steps. One can also confirm this remark by using the information generated by QWalk to plot the total variation distance to both the uniform distribution and the approximate stationary distribution at each time step. The plot would show that the former does not converge to zero while the latter does.

\section{Conclusions}

In this work we initially reviewed the basics of the discrete quantum walks in infinite two-dimensional lattices with the possibility of broken links and two different shift operators. The one-dimensional walk is a particular case of the one described. We presented then the QWalk simulator and described its usage. The QWalk is a quantum walk simulator for one- and two-dimensional lattices.
We showed by means of examples how QWalk can be used to simulate double-slit experiments, walkers in finite lattices, detectors and two different kinds of decoherence. Two kinds of lattices are available for two-dimensional walks and three kinds of lattices for one-dimensional walks.

The simulations presented in this paper correspond to recent works in quantum walks, and have been performed with quite simple commands of the QWalk simulator. Apart from having simulated the available results of literature with great precision, the simulator is also an important tool for researchers to carry out new experiments.

One of the most important potentialities of this simulator is the possibility of studying the walk with different boundaries, by defining permanent broken links appropriately. It also makes possible to investigate the influence of detectors in the quantum walk, and the behaviour of the mixing time in different situations. The simulator can be useful to simulate interesting physical settings, such as the double-slit experiment given as example.

In future versions some improvements may be introduced in QWalk. The sintax of the input file, for instance, would be better if one could use numerical constants instead of typing several digits when declaring custom coins or states. The treatment of the broken links may also be improved in order to simplify the definition of complex boundaries and to allow, for instance, the definition of time-dependent broken links, which would allow moving boundary conditions. It may also be useful for the researcher that the simulator saves output files with the wave-function in intermediate instants of the simulation. It should also be interesting to measure the amount of entanglement during the walk.

\paragraph*{Acknowledgements.} The authors thank Amanda Oliveira, Gonzalo Abal and Raul Donangelo for important discussions. FLM also thanks CNPq (Brazilian scientific council) for financial support.

\bibliographystyle{cpc}
\bibliography{qwalk-cpc}

\appendix
\section{Further options available in the simulator}
\label{sec:further}

In this Appendix we describe the options that were not used in the examples. The \verb'CHECK' keyword requests that some consistency checks be carried out during simulation. A sub-option is expected after this keyword: the \verb'STATEPROB' sub-option declares that the unitarity of the state should be checked at each step of simulation; and the \verb'SYMMETRY' sub-option, that the symmetry of the probability distribution is to be checked. Naturally, the latter sub-option should only be used when the symmetry of the probability distribution is suspected \emph{a priori}, in order to confirm that suspicion. In two-dimensional simulations we may select two kinds of symmetry checking. The \verb'XSYMMETRY' sub-option checks the symmetry around $x$ axis, i.e., checks if the probability at site $(x,y)$ is the same as the probability at site $(-x,y)$, for all $x,y$ in the lattice. Analogously, we have the \verb'YSYMMETRY' sub-options.

In order to define a custom coin we must have a \verb'COIN CUSTOM' command in the main section of the input file and must describe the coin on a separated section in the same file. This section must be delimited by \verb'BEGINCOIN' and \verb'ENDCOIN' keywords, and must contain all the entries of the coin matrix, from left to right and from top to bottom, with real and imaginary parts separated by a blank space. For instance, the code
\begin{center}
%\footnotesize
\begin{minipage}{0.6\textwidth}
\begin{verbatim}
BEGINCOIN
 0.707106781186 0.0    0.0 0.707106781186
 0.0 0.707106781186    0.707106781186 0.0
ENDCOIN
\end{verbatim}
\end{minipage}
\end{center}
defines, in one-dimensional walks, the coin
\begin{equation}
C = \frac{1}{\sqrt{2}}
\left( 
\begin{array}{c c}
1&i\\i&1
\end{array}
\right).
\end{equation}
In two-dimensional simulations the description of the coin is analogous.

In order to define a custom state we must have a \verb'STATE CUSTOM' command in the main section of the input file and must describe the coin on a separated section in the same file. This section must be delimited by \verb'BEGINSTATE' and \verb'ENDSTATE' keywords, and must contain all the non-zero amplitudes of the initial state, with real and imaginary parts separated by a blank space. In one-dimensional simulations each of these amplitudes must be preceded by two integers: the first indicating the coin and the second indicating the position of the walker in that basis state. For instance, the code
\begin{center}
%\footnotesize
\begin{minipage}{0.6\textwidth}
\begin{verbatim}
BEGINSTATE
 1 0 0.0 1.0
ENDSTATE
\end{verbatim}
\end{minipage}
\end{center}
defines, in one-dimensional walks, the initial state
\begin{equation}
\ket{ \Psi_0 } = i \ket{ 1 } \ket{ 0 }.
\end{equation}
In two-dimensional simulations the description of the state is analogous except that it requires two integers for the coin and two integers for the position of the walker.

The \verb'SEED' keyword manually sets a random seed for the random number generator. The default is a number obtained from the system clock and usually the user should not change this.

In two-dimensional walks the \verb'AFTERMEASURE' keyword defines the number of iterations that will be carried out after a non-trivial measurement, i.e., after the result of a measurement corresponds to one of the detectors instead of its complement.

\verb'LATTEXTRA' keyword is used quite rarely. It defines an extra space to be reserved for the lattice in order to avoid access to invalid regions of memory during simulation. Its default value is 1 and normally this value should not be changed by the user. When this keyword is used together with \verb'LATTSIZE' and \verb'STEPS' there is a correct order that must be respected: first \verb'LATTEXTRA', then \verb'STEPS' and finally \verb'LATTSIZE'. Using \verb'LATTEXTRA' keyword may cause QWalk to provide bad gnuplot scripts. In this case the user should fix the script, providing appropriate ranges and ``tics''.

%
% Sometimes program descriptions have Test Run Output at the end.
%
%\newpage
%\hspace{1pc}
%{\bf TEST RUN OUTPUT}
%
%\bigskip
%If your Test Run Output material has to be reproduced as camera ready copy,
%then you can simply paste it under this heading. But you can also list it
%using the verbatim environment of \LaTeX. If you only give short
%pieces of Test Run Output, you can also include these in the Long
%Write-Up of the paper, using the verbatim environment.
%

\end{document}